\documentclass{uncecomp2015}

\usepackage{bstnotations}
\usepackage{amsmath}
\usepackage{amssymb}
\usepackage{amsfonts}
\usepackage{amsbsy}
\usepackage{bbm}

\usepackage{bm}
\newcommand{\norm}[2]{|| #1 ||_{#2}}
\renewcommand{\ve}[1]{\bm{#1}}

\newcommand{\veX}{\ve{X}}

\newcommand{\vea}{\ve{\alpha}}
\newcommand{\veb}{\ve{\beta}}
\newcommand{\veg}{\ve{\gamma}}

\usepackage[FIGBOTCAP]{subfigure}

\usepackage{setspace}

\title{Hierarchical adaptive polynomial chaos expansions}

\author{Chu V. Mai$^1$, Bruno Sudret$^1$}

\heading{Chu V. Mai, Bruno Sudret}

\address{$^1$Chair of Risk, Safety and Uncertainty Quantification, ETH Z\"{u}rich \\
  Stefano-Franscini-Platz 5, 8093 Zurich, Switzerland\\
  e-mail: mai@ibk.baug.ethz.ch, sudret@ibk.baug.ethz.ch}


\abstract{Polynomial chaos expansions (PCE) are widely used in the framework of uncertainty quantification. However, when dealing with high dimensional complex problems, challenging issues need to be faced. For instance, high-order polynomials may be required, which leads to a large polynomial basis whereas usually only a few of the basis functions are in fact significant. Taking into account the sparse structure of the model, advanced techniques such as sparse PCE (SPCE), have been recently proposed to alleviate the computational issue. In this paper, we propose a novel approach to SPCE, which allows one to exploit the model's hierarchical structure. The proposed approach is based on the adaptive enrichment of the polynomial basis using the so-called principle of heredity. As a result, one can reduce the computational burden related to a large pre-defined  candidate set while obtaining higher accuracy with the same computational budget.
}

\begin{document}
\section{INTRODUCTION}

Nowadays, uncertainty quantification (UQ) has become a key topic in computational science and engineering applications. Its formulation comprises uncertainty propagation, which deals with propagating the uncertainties in the input parameters of a computational model to the output quantities of interest. Typical applications include, \eg conducting reliability analysis of complex systems, estimating the statistical distribution of the outputs, computing sensitivity indices, \etc

Uncertainty propagation is traditionally conducted by means of sampling-based methods such as Monte Carlo simulation, importance sampling and subset simulation. These methods, however, are often not suitable for practical problems because of the computational cost related to evaluating complex models repeatedly to obtain a sufficient accuracy. For this reason, inexpensive-to-evaluate surrogate models have been of particular interest as substitutes of the full models. One way of building surrogate models is by means of the spectral approach, in which the quantity of interest (QoI) is considered as a function of the input parameters in a suitable functional space and then projected onto an appropriate basis. Multiple basis choices are available in the literature, including trigonometric, radial basis, wavelet and polynomial functions. In particular, the use of polynomial chaos expansions (PCE) has been found highly effective in numerous applications, see \eg \cite{Xiu2003, Paffrath2007, Young2013, Vaz2014a}.

Building a polynomial chaos expansion requires determining the coefficients corresponding to the polynomial chaos basis. For black-box type problems, it is impossible to directly estimate the polynomial chaos coefficients by means of intrusive approaches that require prior knowledge of the governing equations of the systems. In such cases, one has to rely on non-intrusive methods such as projection, regression which require only a finite number of observations of the system. Among those, least-square minimization methods have received particular interest \cite{Choi2004a, Berveiller2006a}. However, they are facing the so-called \emph{curse-of-dimensionality}, \ie the computational cost increases exponentially with the dimension of the input space. This is the case, \eg when a large number of basis elements is involved but only a small number of experiments (observations) is available, which is not sufficient for the accurate estimation of the coefficients. Several numerical methods have been proposed recently for alleviating such issue, including compressive sensing \cite{Doostan2011}, $l_2$-norm regularized regression \cite{Fagiano2012} and least angle regression \cite{BlatmanJCP2011}. Most of the existing methods rely on considering a set of candidate polynomials that is selected \emph{a priori}, then computing the corresponding coefficients either by setting up a constrained optimization problem or by making assumptions on the structure of the system. Recently, \cite{Perko2014} and \cite{Jakeman2014} proved that in some cases it can be more effective to apply an adaptive basis PCE approach instead. In this paper, we introduce a novel hierarchical method that consists in updating the candidate set adaptively instead of fixing it.

The paper is organized as follows: in Section 2, we recall briefly the non-intrusive regression-based approach for computing PCE. In Section 3, the hierarchical adaptive method for PCE is introduced. For the sake of illustration, we prove the effectiveness of the proposed method on two numerical examples.

\section{POLYNOMIAL CHAOS EXPANSIONS}

\subsection{Polynomial chaos expansions}
Consider a system with uncertain input parameters whose behavior can be described by the following equation:
\begin{eqnarray}
Y = \cm(\veX),
\label{eq2.1.1}
\end{eqnarray}
in which $\veX=\acc{X_1 \enum X_M}\tr$ is a vector of $M$ independent uncertain input parameters and $Y$ is the output quantity of interest. For the sake of simplicity, and without loss of generality, one considers only the case of a scalar-valued output.
Note that one does not know Eq.~\eqref{eq2.1.1} explicitly when the computational model is of a black-box type. 


In the Hilbert space of square-integrable functions of variables $\acc{X_i,  i=1 \enum M}\tr$, one can select a basis of orthonormal univariate polynomials ${\{ \psi_k^i, k \in \Nn \} }$, also known as polynomial chaos functions \cite{Soize2004,Xiu2002}, associated with the probability density measure ${ \mathbb{P}_{X_i}(\di x_i)= f_{X_i}(x_i) \di x_i }$. For instance when $X_i$ is a uniform (resp. standard normal) random variable, the corresponding basis consists of orthonormal Legendre (resp. Hermite) polynomials. The polynomial chaos expansion of $Y$ reads:
\begin{eqnarray}
 Y= \sum\limits_{\vea \in \Nn^{M}} y_{\vea} \ve{\psi}_{\vea}(\veX)
 \label{eq2.2.1}
\end{eqnarray}
where $\vea = (\alpha_1 \enum \alpha_M)$ is a multi-index with $\alpha_i, i=1\enum M$ denoting the degree of the polynomial in the direction of $X_i$, $\ve{\psi}_{\vea}(\veX) = \prod\limits_{i}^{M} \psi_{\alpha_i}^i(X_i)$ are multivariate polynomials obtained by the tensor product of univariate polynomials and $y_{\vea}$ are the associated coefficients.

The regression-based adaptive sparse polynomial chaos expansions (SPCE) consists of two sequential steps. First, a set of candidate polynomial chaos basis is chosen \emph{a priori} (see Section~\ref{sec2.3}). Second, the relevant basis are selected from the candidate set and the corresponding coefficients are computed (see Section~\ref{sec2.4}). The two steps are described subsequently.

\subsection{Truncation schemes}
\label{sec2.3}
In practice, an approximate form of PCE in Eq.~\eqref{eq2.2.1} with a finite number of terms must be used: 
\begin{eqnarray}
 Y \approx \sum\limits_{\vea \in \ca} y_{\vea} {\psi}_{\vea}(\veX)
 \label{eq2.3.1}
\end{eqnarray}
The set of multi-indices $\ca $ is obtained by truncating the $\Nn^{M}$ space. We present herein two truncation schemes that were recently proposed in \cite{BlatmanThesis}.

In the first scheme, one assumes that the main effects and the low-degree interaction effects are more important than high-degree interaction effects. Thus, in the multi-indices space, the relevant terms lie in the subspace defined as follows \cite{BlatmanThesis}:
\begin{eqnarray}
	 \ca \equiv \ca_q^{M,p} =\{ \vea \in \Nn^M: \norm{\vea}{q} =  \prt{\sum\limits_{i=1}^M \alpha_i ^{q}}^{1/q}   \leqslant p \,  \}
	 \label{eq2.3.2}
\end{eqnarray}
where $0<q<1$ is the parameter governing the \emph{hyperbolic truncation scheme} and $p$ is the prescribed maximum degree of the polynomials.
Note that in the special case when $q$ is assigned the value of 1, the standard truncation scheme is recovered:
\begin{eqnarray}
\ca \equiv \ca^{M,p} =\{ \vea \in \Nn^M: \norm{\vea}{1} =  \sum\limits_{i=1}^M \alpha_i   \leqslant p \}
\label{eq2.3.3}
\end{eqnarray}
Given a prescribed value of $p$, the size of the polynomial basis obtained with the hyperbolic truncation scheme increases as $q$ increases and attains its maxima at $q=1$.

In the second scheme, one assumes that low-rank effects are more important than high-rank effects, \ie two-dimensional interaction terms are more relevant than three-dimensional interaction terms and so on. This assumption is based on the so-called \emph{hierarchy principle} \cite{Yuan2007} which states that the model can be approximated by low-rank terms. The \emph{low-rank truncation scheme} reads \cite{BlatmanThesis}: 
\begin{eqnarray}
	 \ca \equiv \ca_r^{M,p} =\{ \vea \in \Nn^M: \norm{\vea}{0} = \sum\limits_{i=1}^M \mathbbm{1}_{\alpha_i > 0} \leq r, \, \norm{\vea}{1} =  \sum\limits_{i=1}^M \alpha_i   \leqslant p \}
	 \label{eq2.3.4}
\end{eqnarray}
in which $\norm{\vea}{0}$ is the rank of the multivariate polynomial ${\psi}_{\vea}$, defined as the total number of non-zero component indices $\alpha_i, i= 1 \enum M$. The prescribed rank $r$ is chosen as a small integer value, \eg $r=2,3,4.$

The proposed schemes, namely hyperbolic and low-rank truncation, usually result in a candidate basis of manageable size.
The two strategies consist in selecting a set of candidate basis \emph{a priori}, from which a sparse subset is then retained. The accuracy of the PCE is measured by means of error estimates, \eg empirical error or leave-one-out (LOO) error $\epsilon_{LOO}$ \cite{SudretBook2015}.

\subsection{Least angle regression}
\label{sec2.4}
Given a candidate set generated by a truncation scheme (see Section~\ref{sec2.3}) and an experimental design (a set of input values $\ve{\cx}$ and the corresponding observed output $\cy$ of the model), one has to determine the relevant basis and the associated coefficients of the expansion:
\begin{eqnarray}
 \cy \approx \sum\limits_{\vea \in \ca} y_{\vea} {\psi}_{\vea}(\ve{\cx})
 \label{eq2.4.1}
\end{eqnarray}
This is an example of the variable selection problem in statistics. To this purpose, a wide variety of methods is available, including ordinary least-squares, compressive sensing and least-angle-regression.

\begin{figure}[ht]
  \begin{center}
  \subfigure
  [Candidate set]
  	{
    \includegraphics[width=0.3\linewidth]{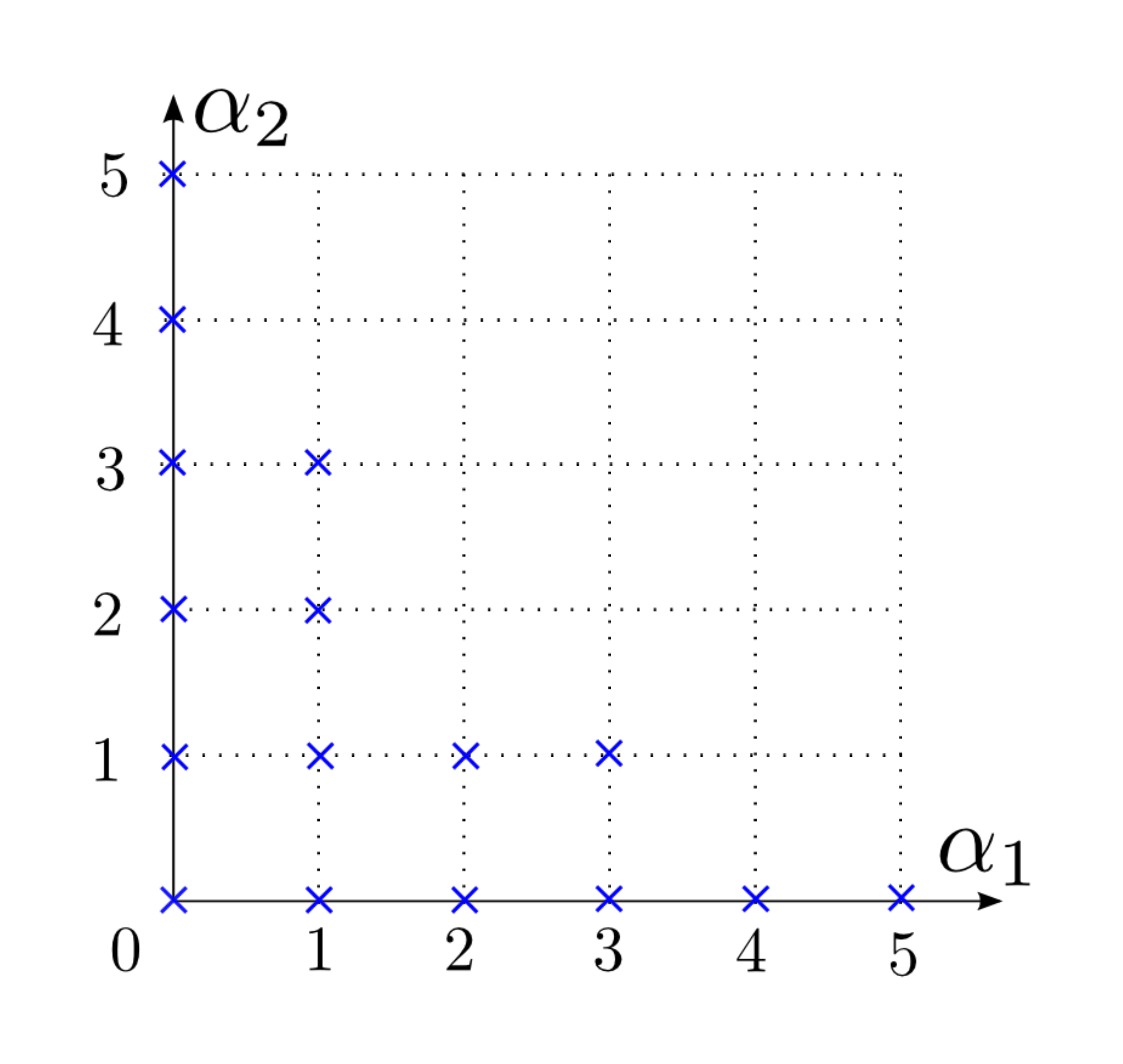}
    	}
    \subfigure
    [Iteration 1]
    	{
    \includegraphics[width=0.3\linewidth]{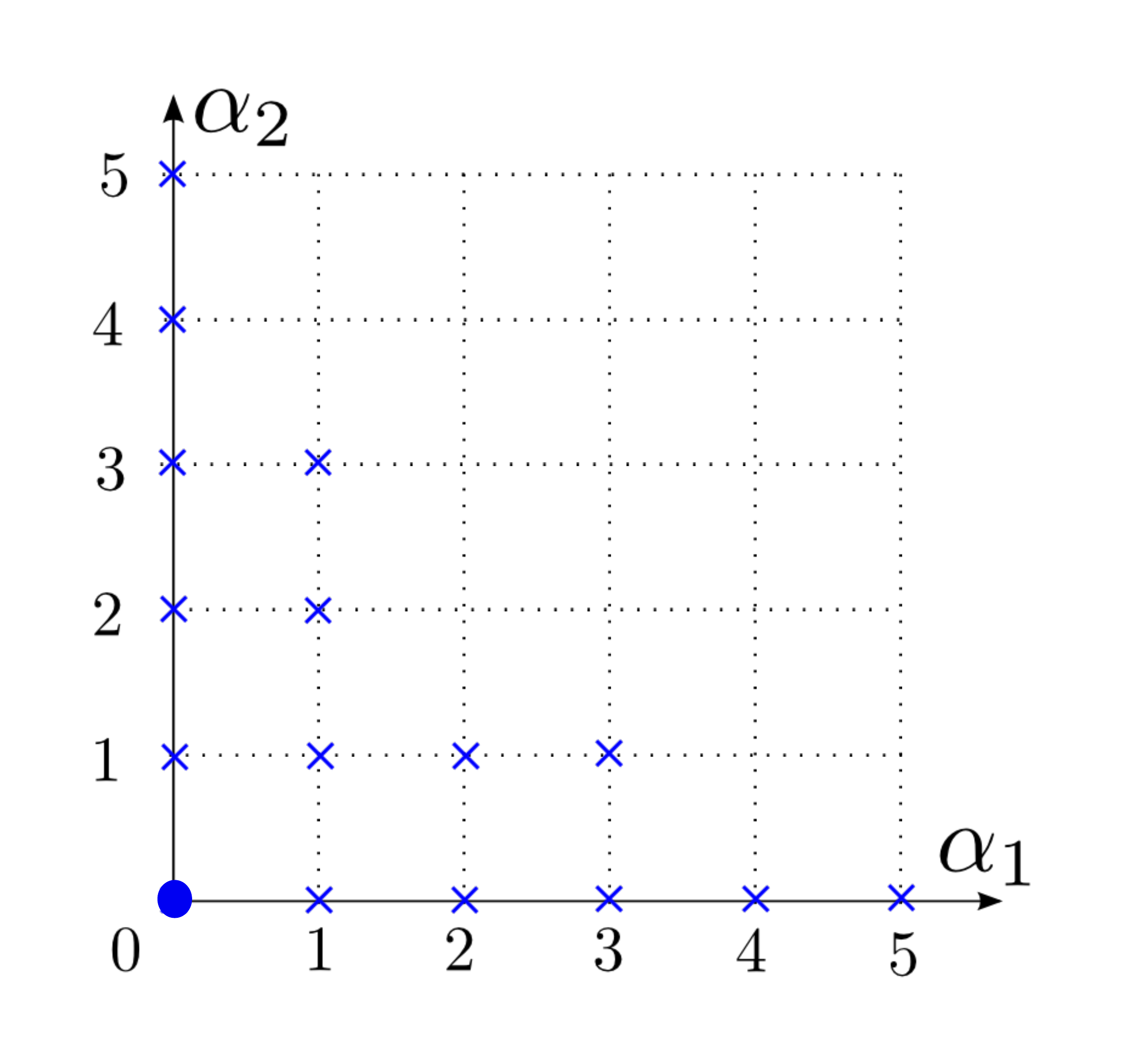}
    	}
    \subfigure
    [Iteration 2]
    	{
    \includegraphics[width=0.3\linewidth]{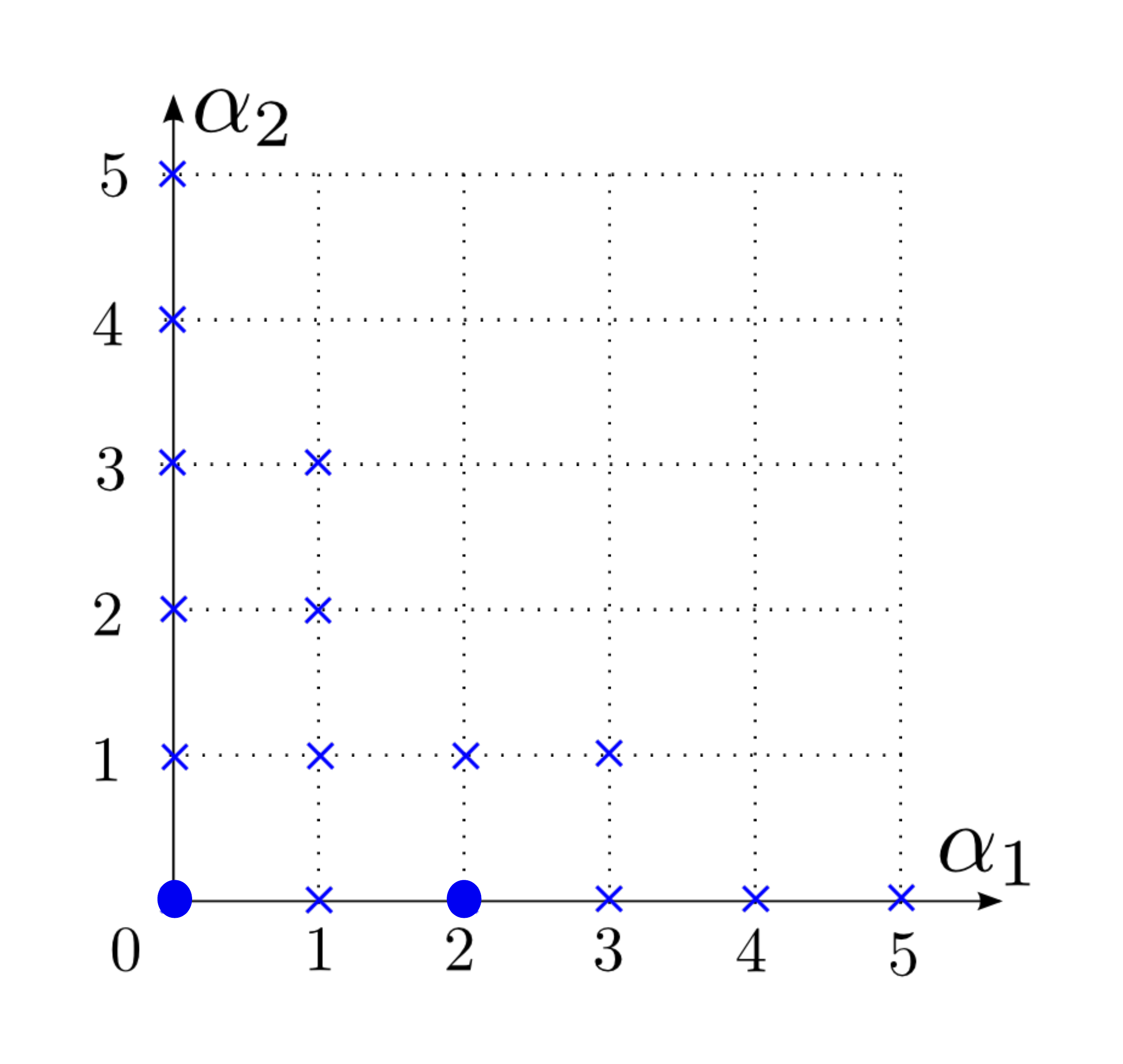}
    	}
    \subfigure
    [Iteration 3]
    	{
    \includegraphics[width=0.3\linewidth]{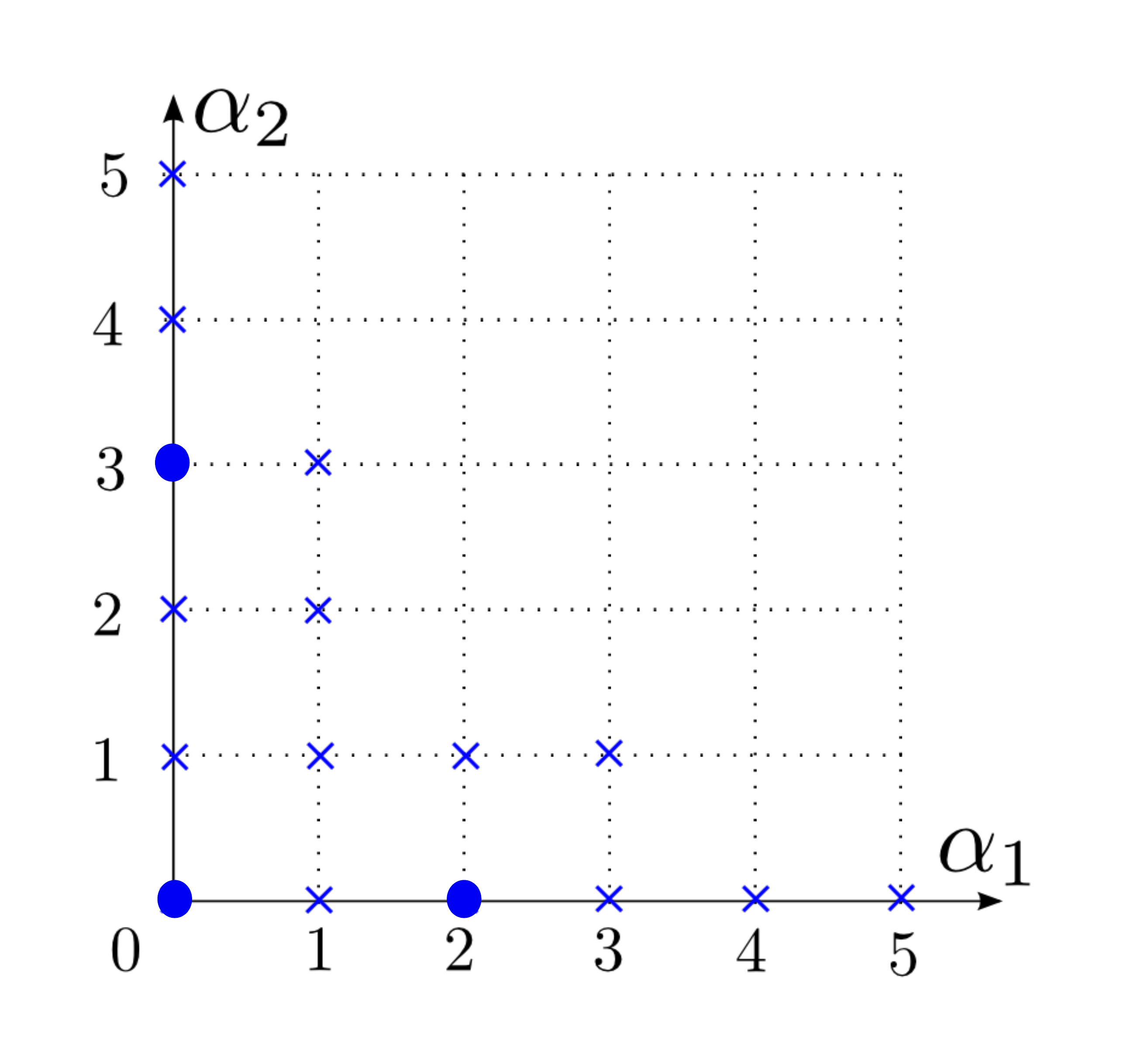}
    	}
    \subfigure
    [Iteration 4]
    	{
    \includegraphics[width=0.3\linewidth]{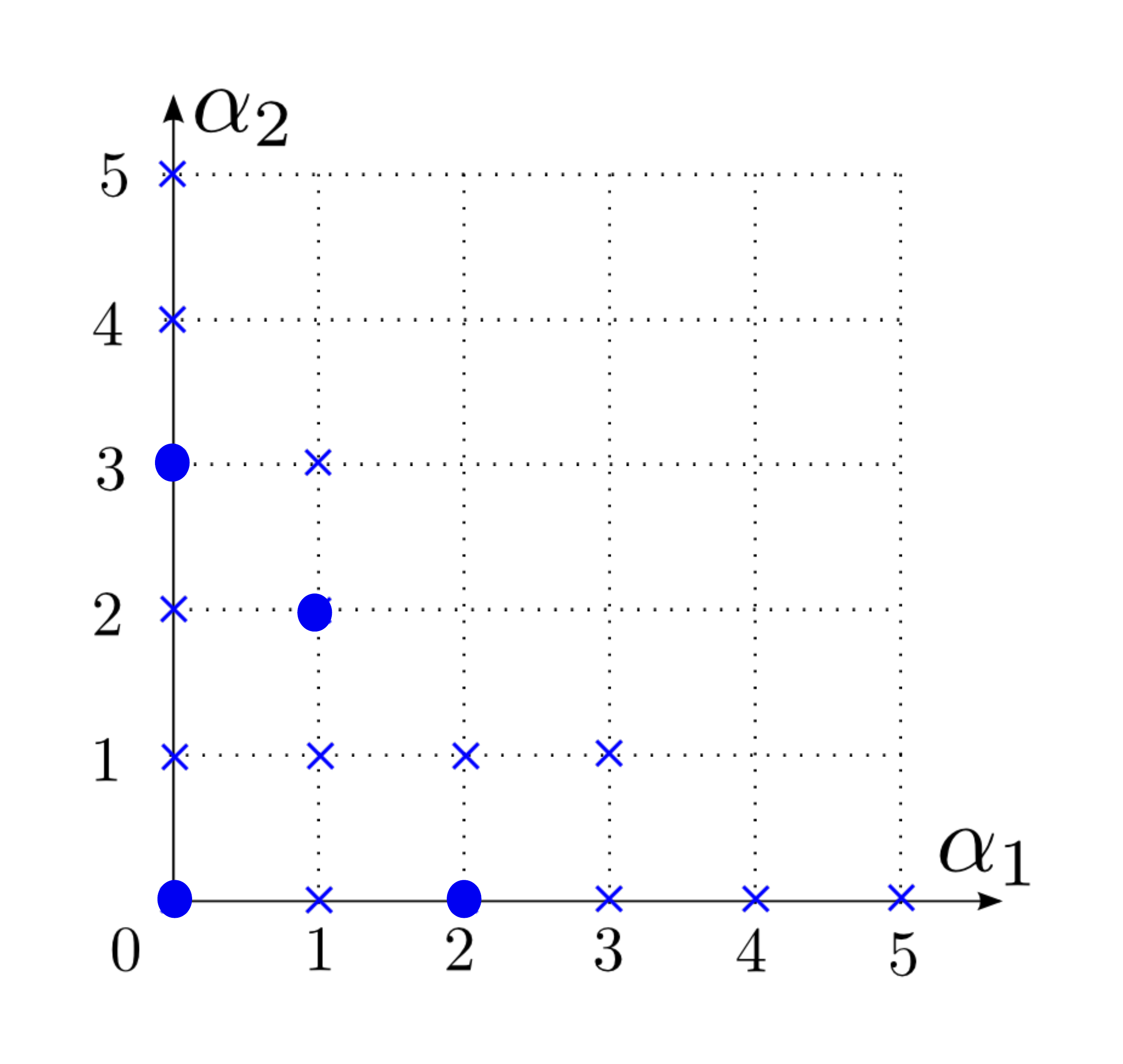}
    	}
    \subfigure
    [Iteration 5]
    	{
    \includegraphics[width=0.3\linewidth]{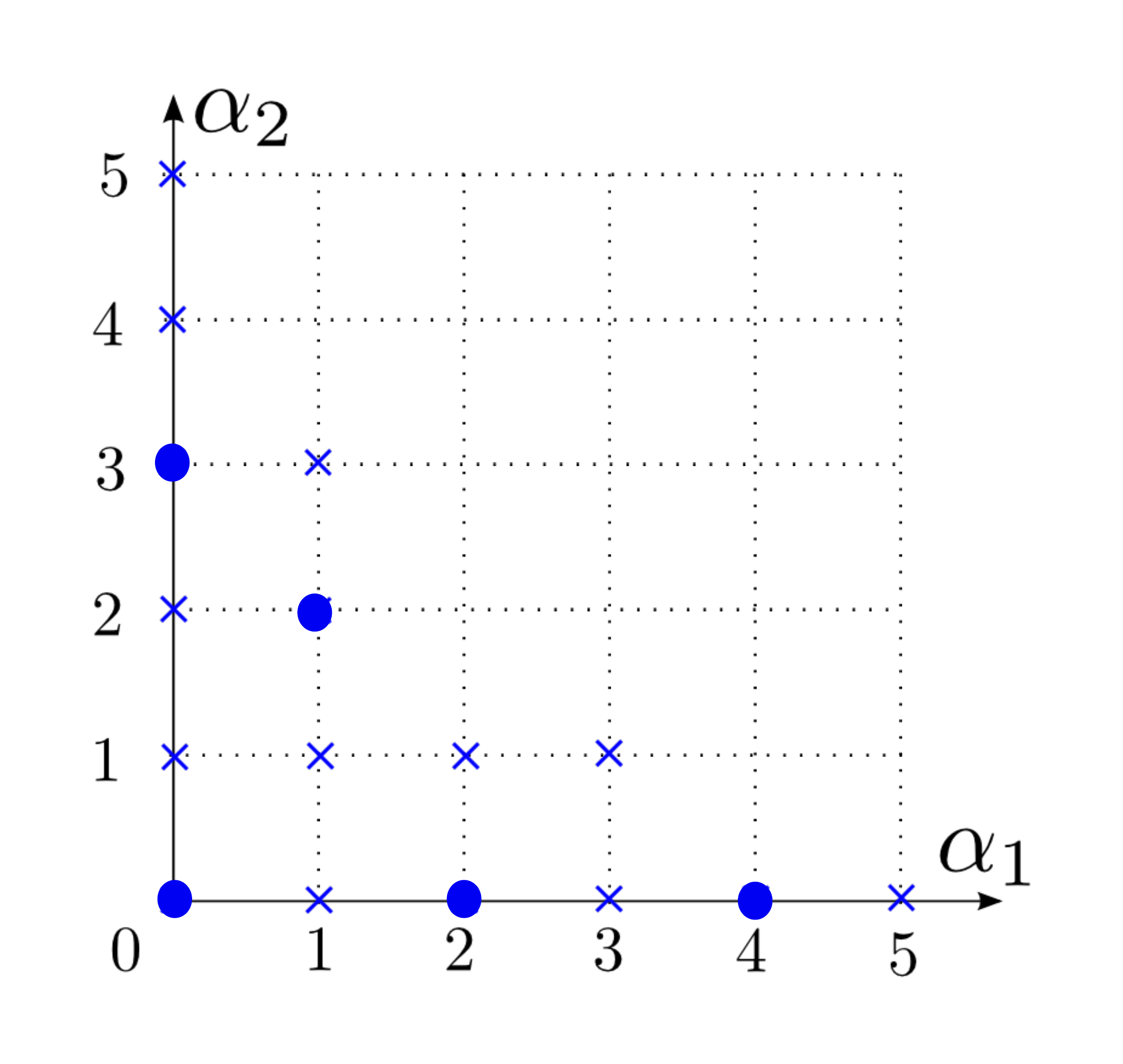}
    	}
    \caption{Least angle regression in the multi-index space. $p=5$, $r=2$ and $q=0.75$ are used for the truncation schemes. The cross symbols (x) represent the candidate basis. The dot symbols ($\bullet$) represent the basis selected by LAR in the current iteration.}
    \label{fig2.4.1}
  \end{center}
  
\end{figure}

Least angle regression (LAR) \cite{Efron2004} is an iterative regression method for variable selection, which was proven to be particularly powerful when the number of regressors is much larger than the number of experiments, and only a few of them are relevant. Assuming that the model of interest respects the \emph{principle of sparsity}, \ie it can be approximated using a relatively small number of polynomials, LAR is used to derive \emph{sparse polynomial chaos expansions} (SPCE) in \cite{BlatmanJCP2011}. 
At each iteration, a basis set is retained with the associated coefficients and a measure of accuracy, \eg the leave-one-out error, is computed. At the end of the process, based on the accuracy measure, one can choose the best performing PCE model. Note that after detecting the relevant basis by LAR, one should recompute the coefficients using ordinary least-square method (hybrid-LAR) \cite{BlatmanThesis}.
The LAR algorithm can be summarized as follows:
%

\begin{enumerate}
	\item Initialize the set of candidate regressors to the full basis and the set of selected regressors to $\varnothing$.
	\item Initialize all coefficients equal to 0. Set the residual equal to the output vector.
	\item For each iteration until the stop condition is satisfied or all the regressors have been analyzed, perform the following steps:
	\begin{itemize}
		\item Determine the most correlated candidate basis element to the current residual and add it to the selected basis
		\item Move simultaneously all the coefficients associated with the selected basis  until another basis element is as correlated with the residual as they are. Update the residual. 
	\end{itemize}
	\item Choose the best iteration based on a prescribed error estimate, \eg $\epsilon_{LOO}$.
\end{enumerate}
Readers are referred to \cite{BlatmanJCP2011} for further details.
The algorithm is schematically illustrated in \figref{fig2.4.1}, in which the adaptive selection of the relevant basis is presented for the first 5 terms in the expansion.
The SPCE based on low-rank hyperbolic truncation and LAR is hereafter considered as the reference approach.

\section{HIERARCHICAL ADAPTIVE POLYNOMIAL CHAOS EXPANSIONS}

Recently, \cite{Perko2014} and \cite{Jakeman2014} proved that adaptive basis scheme can be more efficient than fixing a candidate basis a priori. Moreover, \cite{Perko2014} propose a novel adaptive-basis non-intrusive spectral projection-based approach for PCE. An initial basis is enriched by adding candidate sets associated with adaptive sparse grids and removing the unimportant PC basis elements with minor contributions to the variance. In \cite{Jakeman2014}, it is assumed that a basis function is admissible, or potentially relevant, only if its backwards neighbors exist in every dimension. Given the same computational budget, the authors observe that adaptive basis selection schemes provide more accurate PCE than predefined basis schemes.

In this section, we introduce a novel adaptive basis SPCE scheme which relies on LAR and the so-called \emph{principle of heredity}. We emphasize that the models of interest are also assumed to follow the principle of sparsity.


\subsection{Principle of heredity}

Heredity is a biological process in which the characteristics of parents are passed down to their child through the genes. This process is represented by Mendel's principle of heredity \cite{Bateson2013} which was recently generalized in a statistical sense in \cite{Hamada1992}. According to the principle of heredity, if two factors are not relevant in the model, it is likely that their interaction is also irrelevant. This principle is nowadays used for the detection of important interaction effects (interaction screening) in high and ultra-high dimensional problems.
It was incorporated in variable selection techniques, \eg forward selection scheme \cite{Hao2014}, LAR \cite{Yuan2007} and structured variable selection and estimation \cite{Yuan2009}. 

Herein, we introduce the use of the heredity principle in the context of SPCE.
Consider a the model $Y(\veX)$ with $\veX$ denoting the vector of $M$ independent input parameters. Based on the hierarchy principle, one assumes that $Y$ can be approximated with only main effects and two-dimensional interaction effects as follows:
\begin{eqnarray}
 Y= y_{\ve{0}} + \sum\limits_{\veb \in \ca_{1}} y_{\veb} \ve{\psi}_{\veb}(\veX) + \sum\limits_{\veg \in \ca_{2}} y_{\veg} \ve{\psi}_{\veg}(\veX) + \epsilon
 \label{eq3.1.1}
\end{eqnarray}
where $\ca_{1} =\acc{\veb \in \Nn^M: \norm{\veb}{0} =1, \norm{\veb}{1} \leqslant p}$ and  $\ca_{2} =\acc{\veg \in \Nn^M: \norm{\veg}{0} =2, \norm{\veg}{1} \leqslant p}$, \ie the second (resp. third) summand consists of univariate (resp. bivariate) polynomials; $y_{\ve{0}}$ is the constant term, $y_{\veb}$ and $y_{\veg}$ are the coefficients of the expansion and $\epsilon$ is the truncation residual error.

\begin{figure}[ht]
  \begin{center}
	\subfigure
      [Weak heredity]
      	{
        \includegraphics[width=0.45\linewidth]{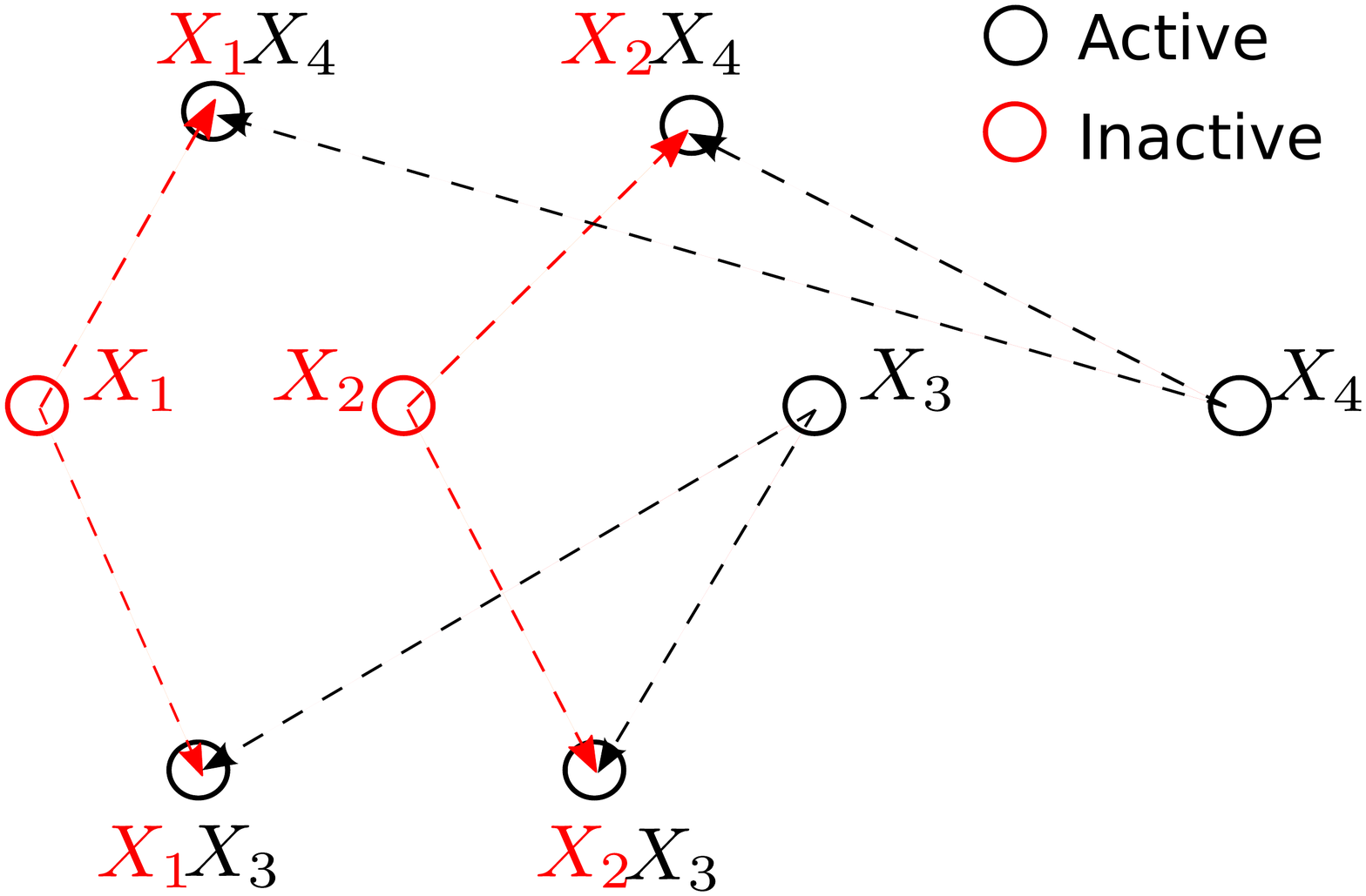}
        }
    \subfigure
      [Strong heredity]
      	{
        \includegraphics[width=0.45\linewidth]{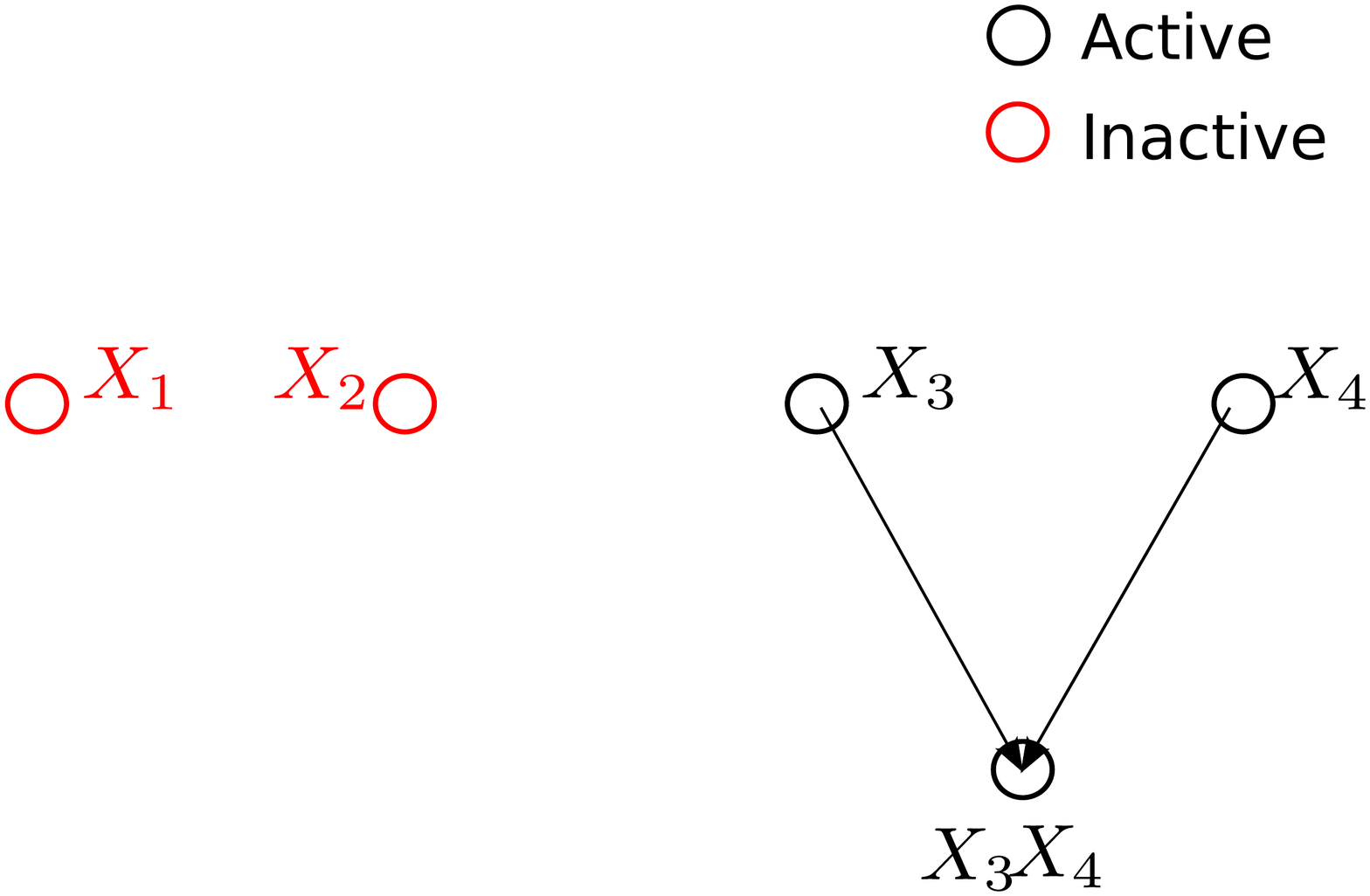}
        }
        \caption{Heredity principle}
        \label{fig3.1.1}
  \end{center}
  
\end{figure}

The heredity principle states that the interaction term $\ve{\psi}_{\veg}(\veX) \equiv \psi_{\gamma_i}^i(X_i) \psi_{\gamma_j}^j(X_j), i\neq j $ is active in the model only if the parent terms $\psi_{\gamma_i}^i(X_i)$ and $\psi_{\gamma_j}^j(X_j)$ are also present \cite{Hao2014}. More precisely, two forms of the heredity principle may be considered, namely weak heredity and strong heredity:
\begin{itemize}
	\item \emph{Weak heredity} indicates that given an active interaction term $\psi_{\gamma_i}^i(X_i) \psi_{\gamma_j}^j(X_j)$, at least one of the two parent terms $\psi_{\gamma_i}^i(X_i)$ and $\psi_{\gamma_j}^j(X_j)$ exist in the model.
	\item \emph{Strong heredity} implies that both parent terms $\psi_{\gamma_i}^i(X_i)$ and $\psi_{\gamma_j}^j(X_j)$ must be active for the interaction term $\psi_{\gamma_i}^i(X_i) \psi_{\gamma_j}^j(X_j)$ to be included in the model.
\end{itemize}
\figref{fig3.1.1} illustrates the principle in the example of a four-dimensional model, which requires only first degree univariate polynomials and their interactions.

To introduce the principle of heredity in SPCE, we need to combine this principle with the LAR method. To this end, \cite{Yuan2007} introduced distinct generalized LAR algorithms for strong and weak heredity. The proposed algorithms, however, cannot solve problems in which both strong and weak heredity are active simultaneously, \eg one group of variables follows strong heredity, another group follows weak heredity. A clear distinction between strong and weak heredity in a practical problem is, however, quasi-impossible a priori. This limitation is overcome by the approach proposed in the next section.

\subsection{Hierarchical adaptive PCE}

\begin{figure}[ht!]
  \begin{center}
  \subfigure
  [Initial candidate]
  	{
    \includegraphics[width=0.3\linewidth]{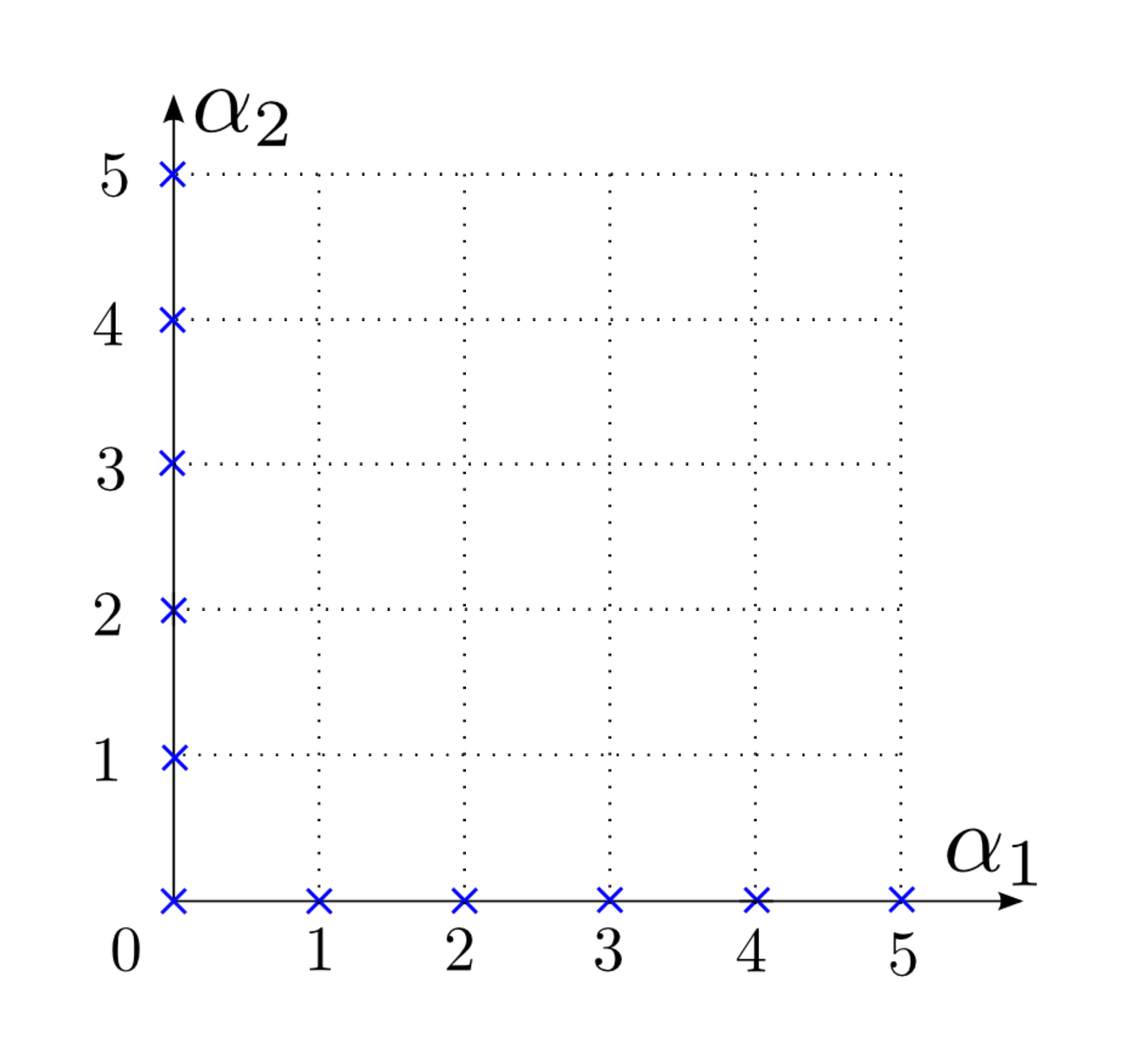}
    }
  \subfigure
    [Iteration 1]
    	{
    \includegraphics[width=0.3\linewidth]{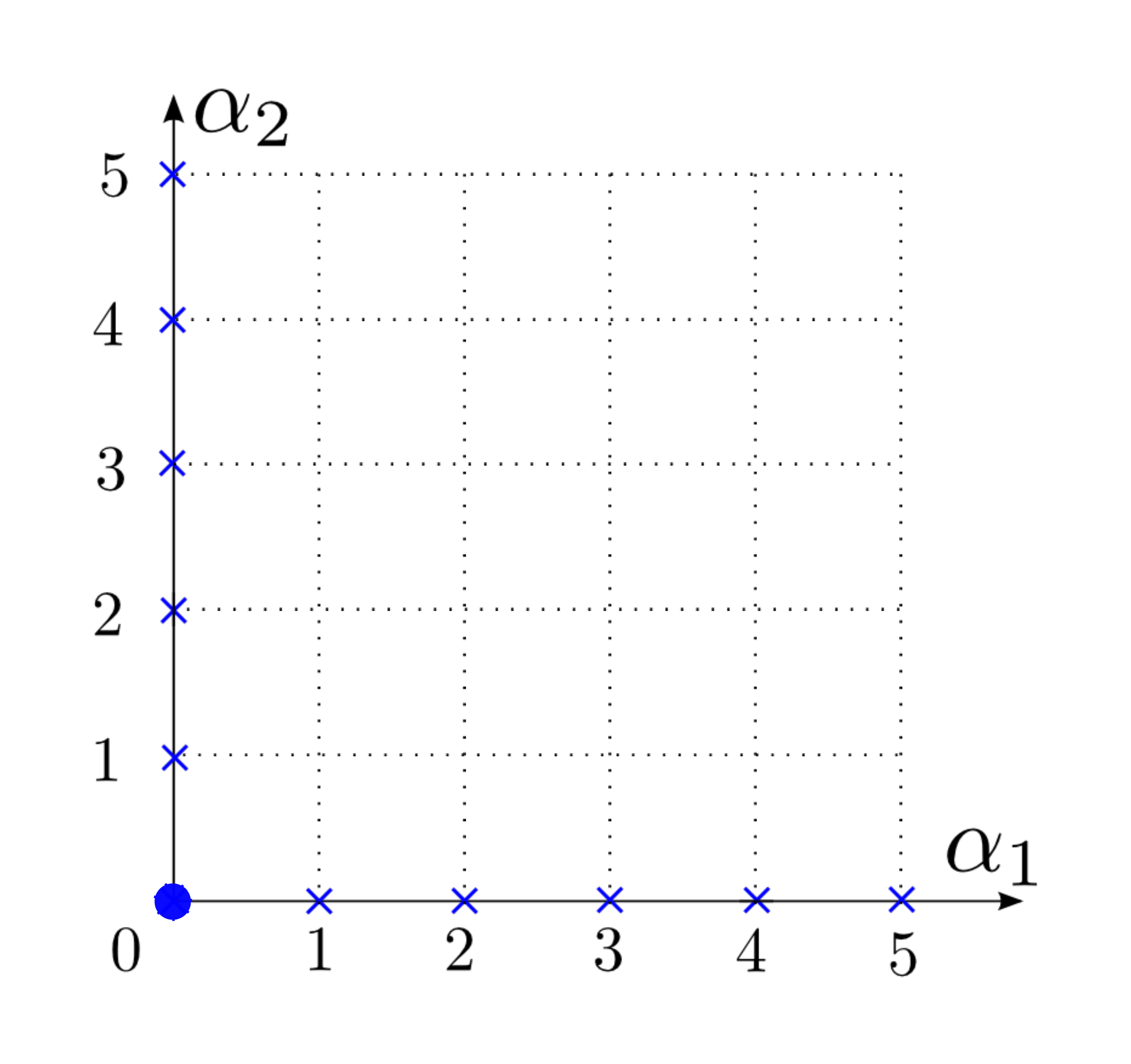}
    }
  \subfigure
  [Iteration 2]
  	{
    \includegraphics[width=0.3\linewidth]{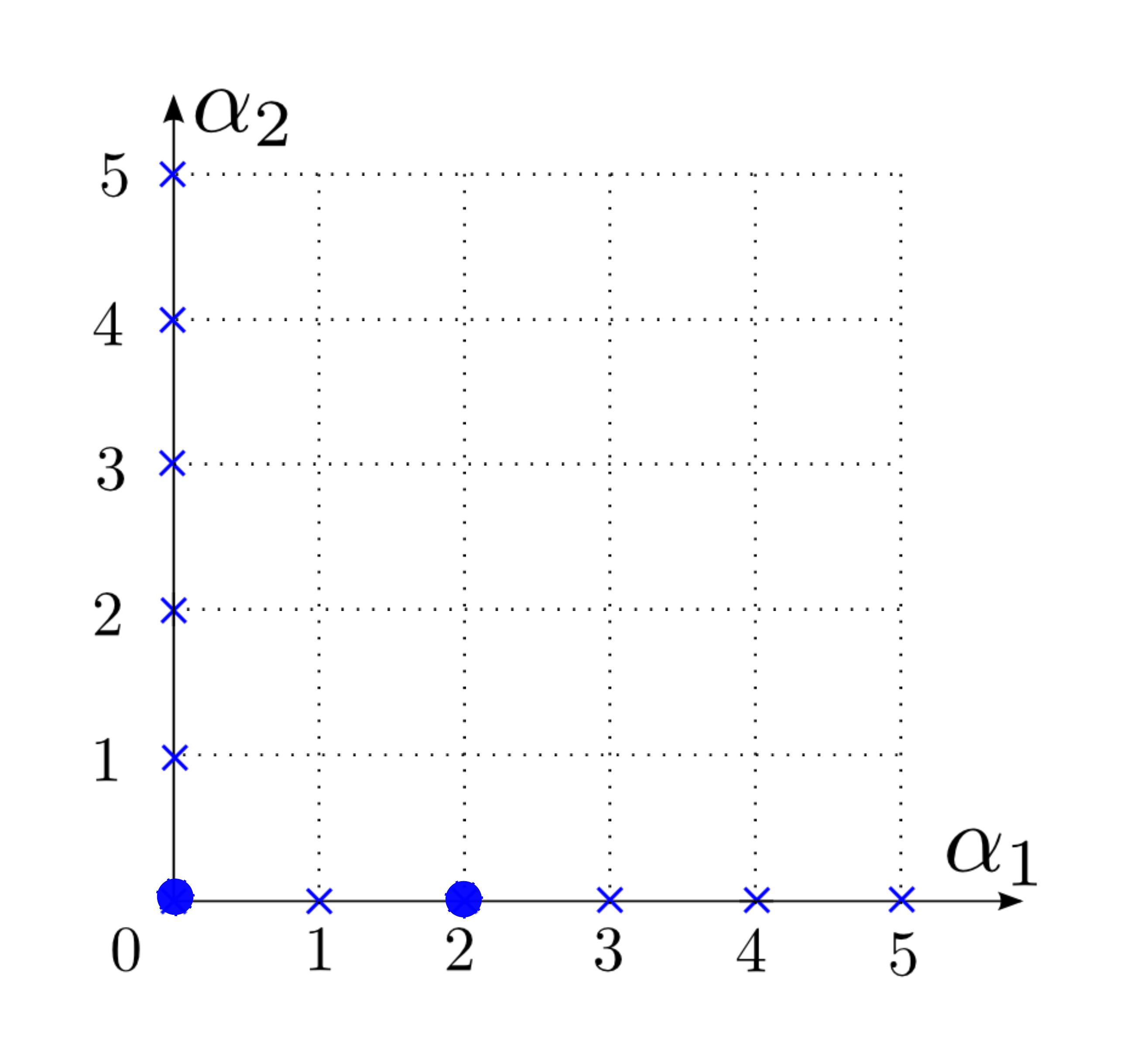}
    }
\subfigure
[Iteration 3]
	{
    \includegraphics[width=0.3\linewidth]{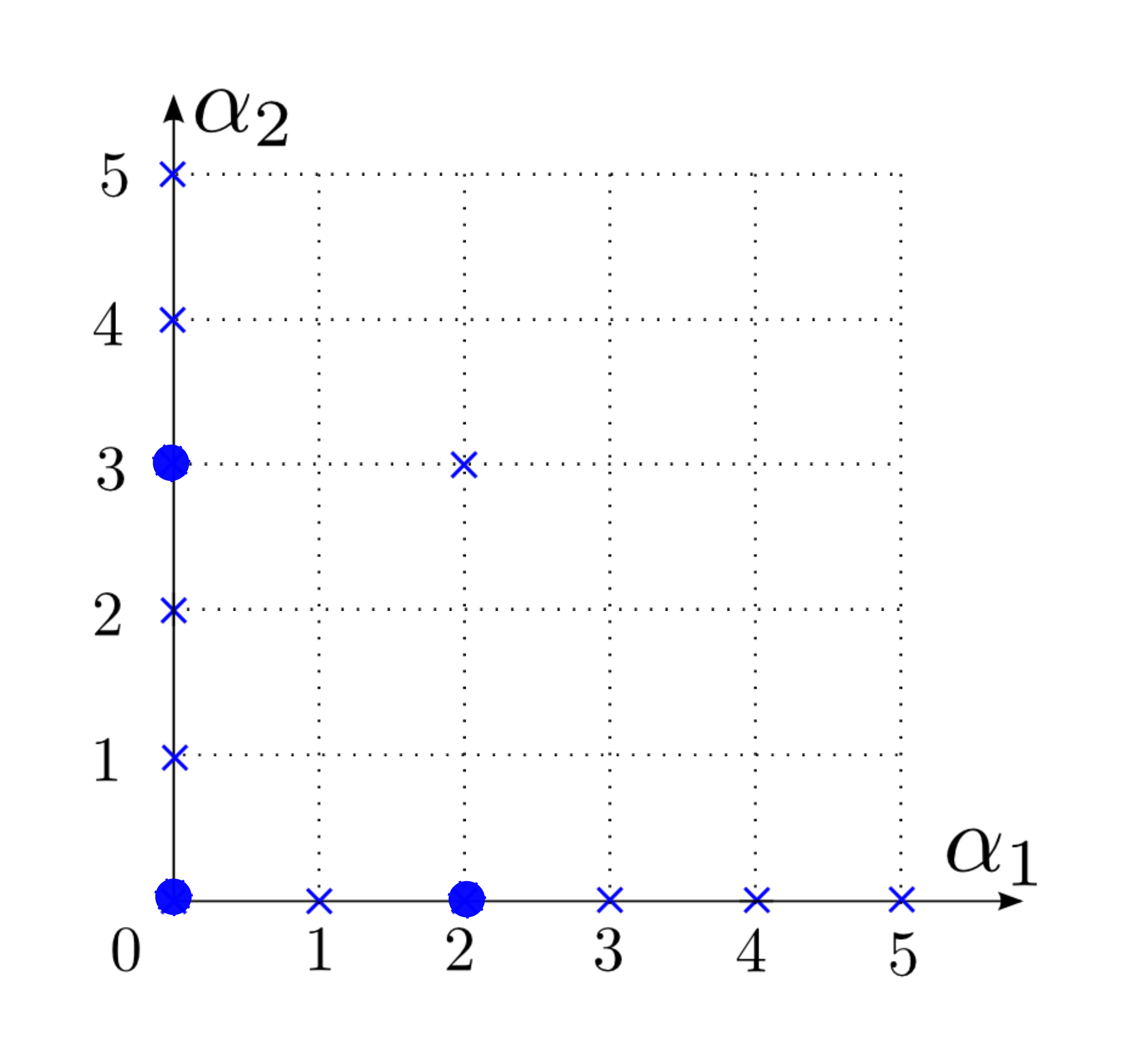}
    }
   \subfigure
   [Iteration 4: strong heredity]
   	{
    \includegraphics[width=0.3\linewidth]{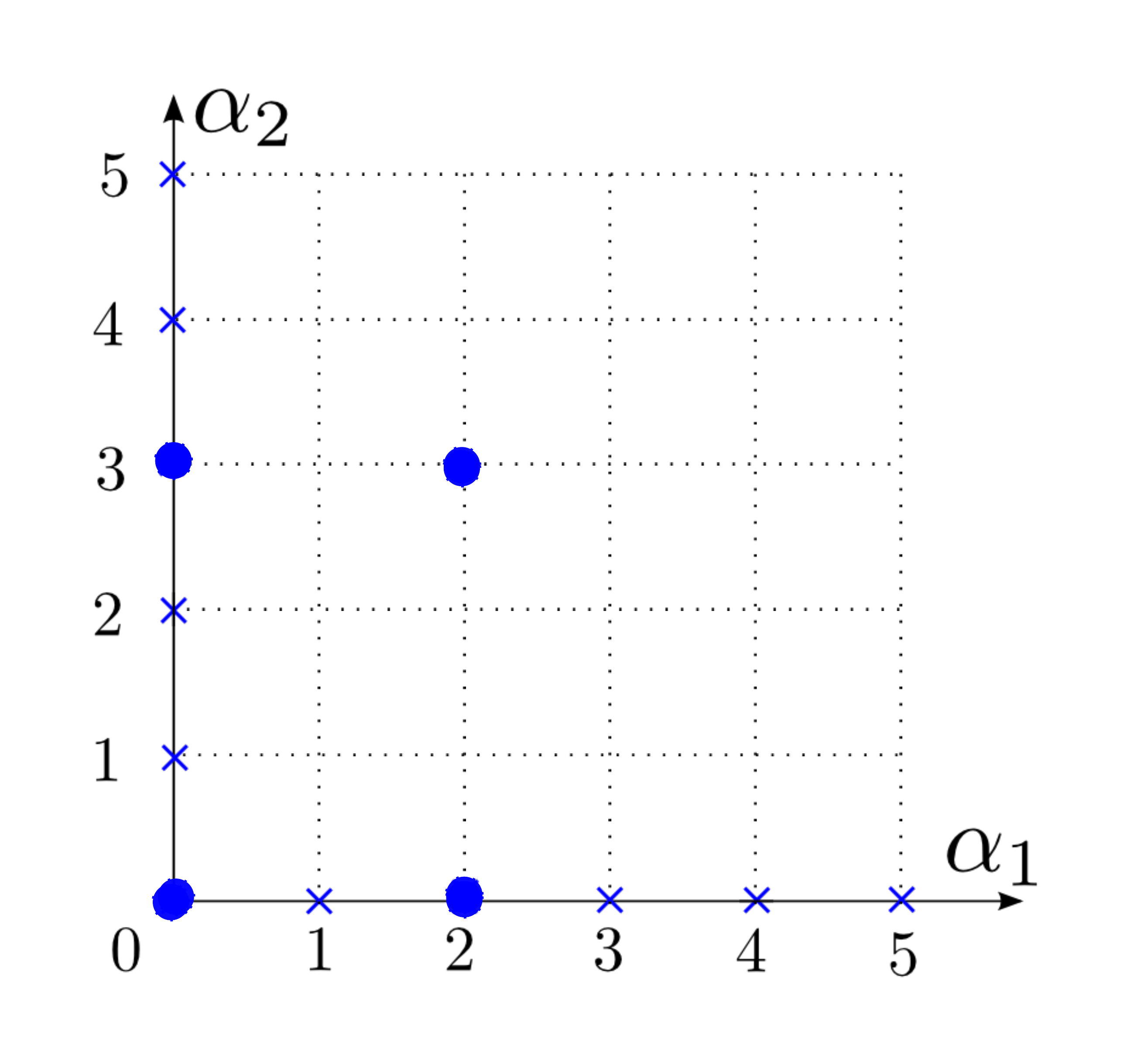}
    }
  \subfigure
  [Iteration 5a]
  	{
    \includegraphics[width=0.3\linewidth]{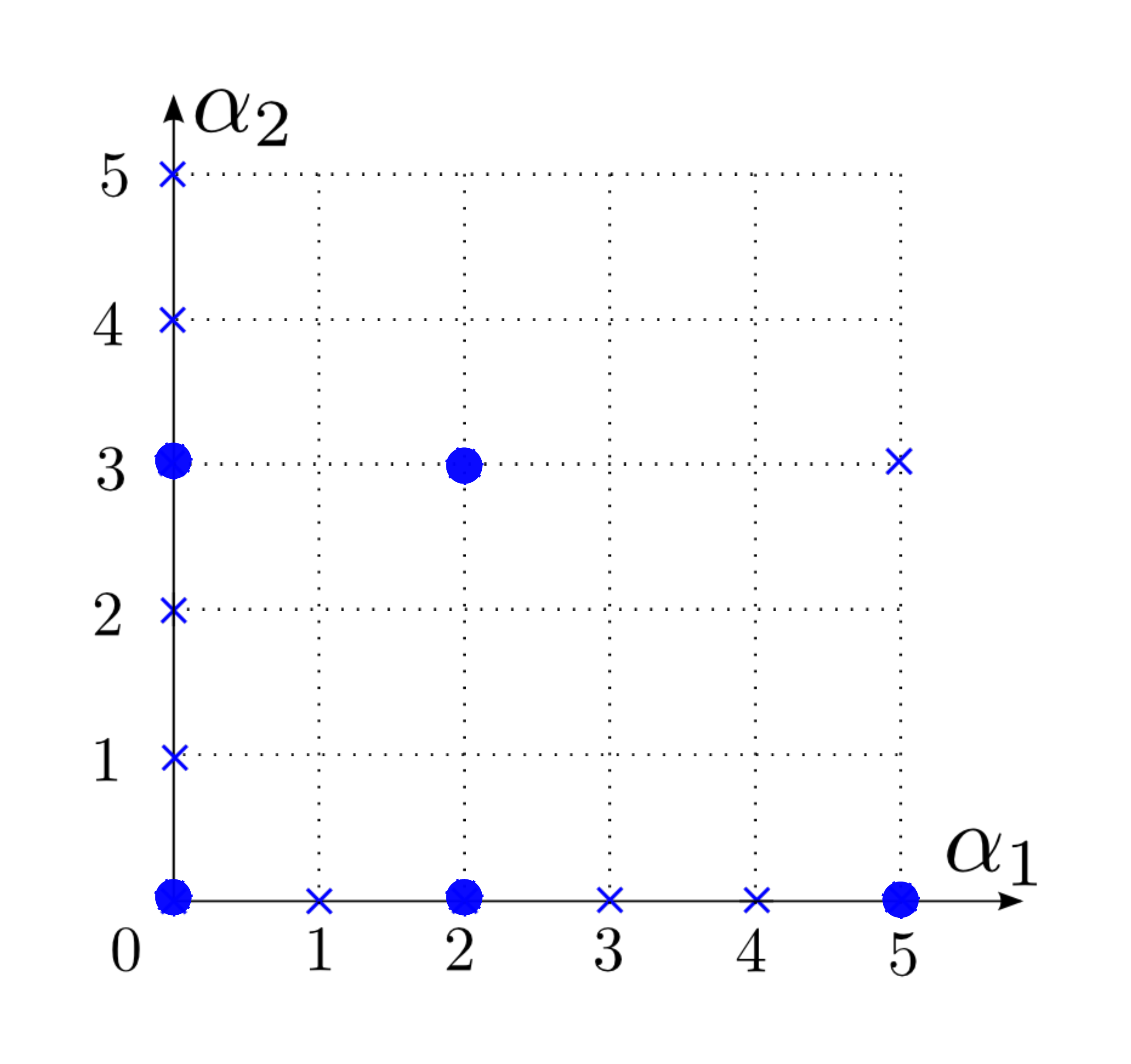}
    }
  \subfigure
  [Iteration 5b: weak heredity]
  	{
    \includegraphics[width=0.3\linewidth]{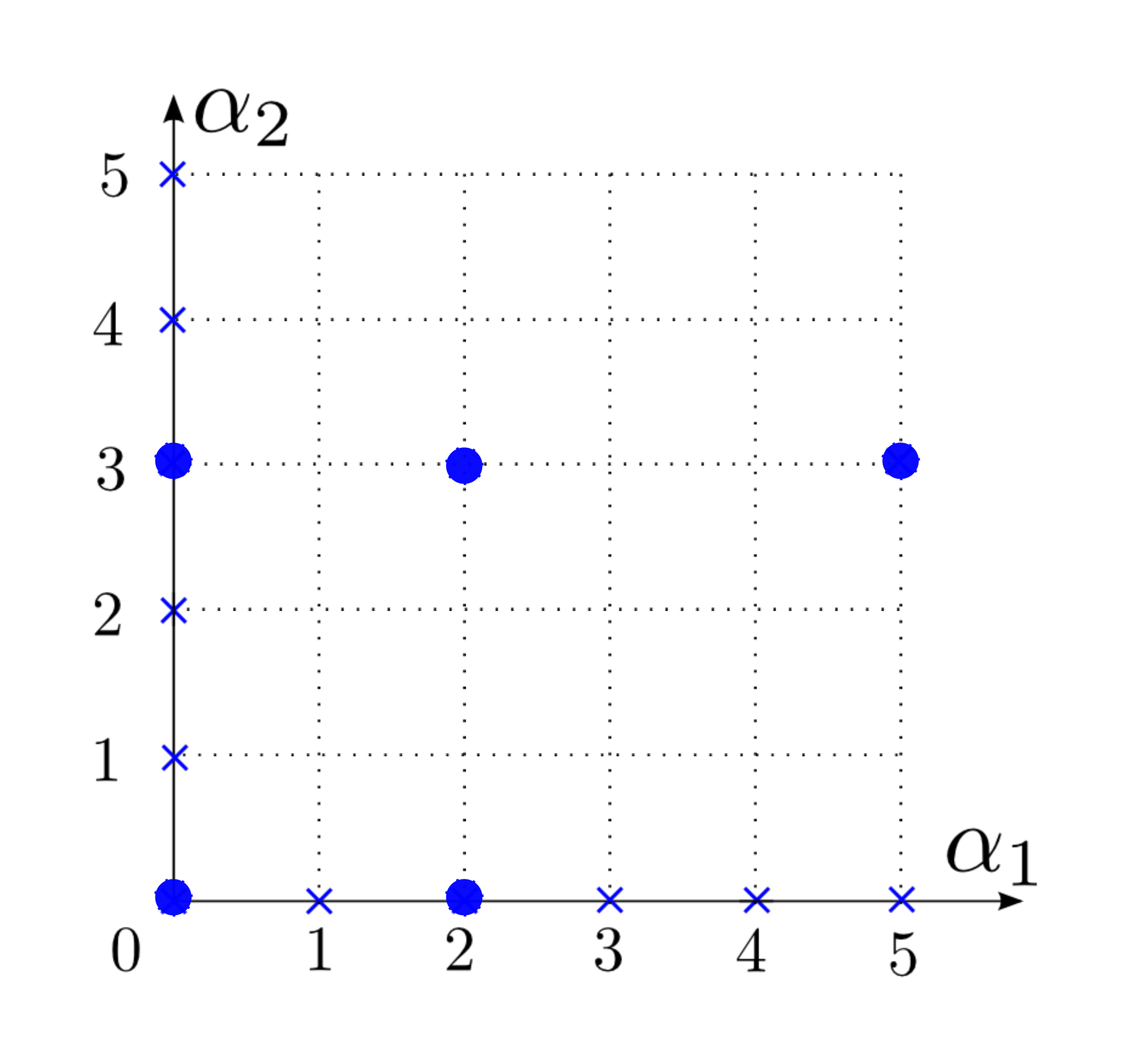}
    }
  \subfigure
  [Iteration 6a]
  	{
    \includegraphics[width=0.3\linewidth]{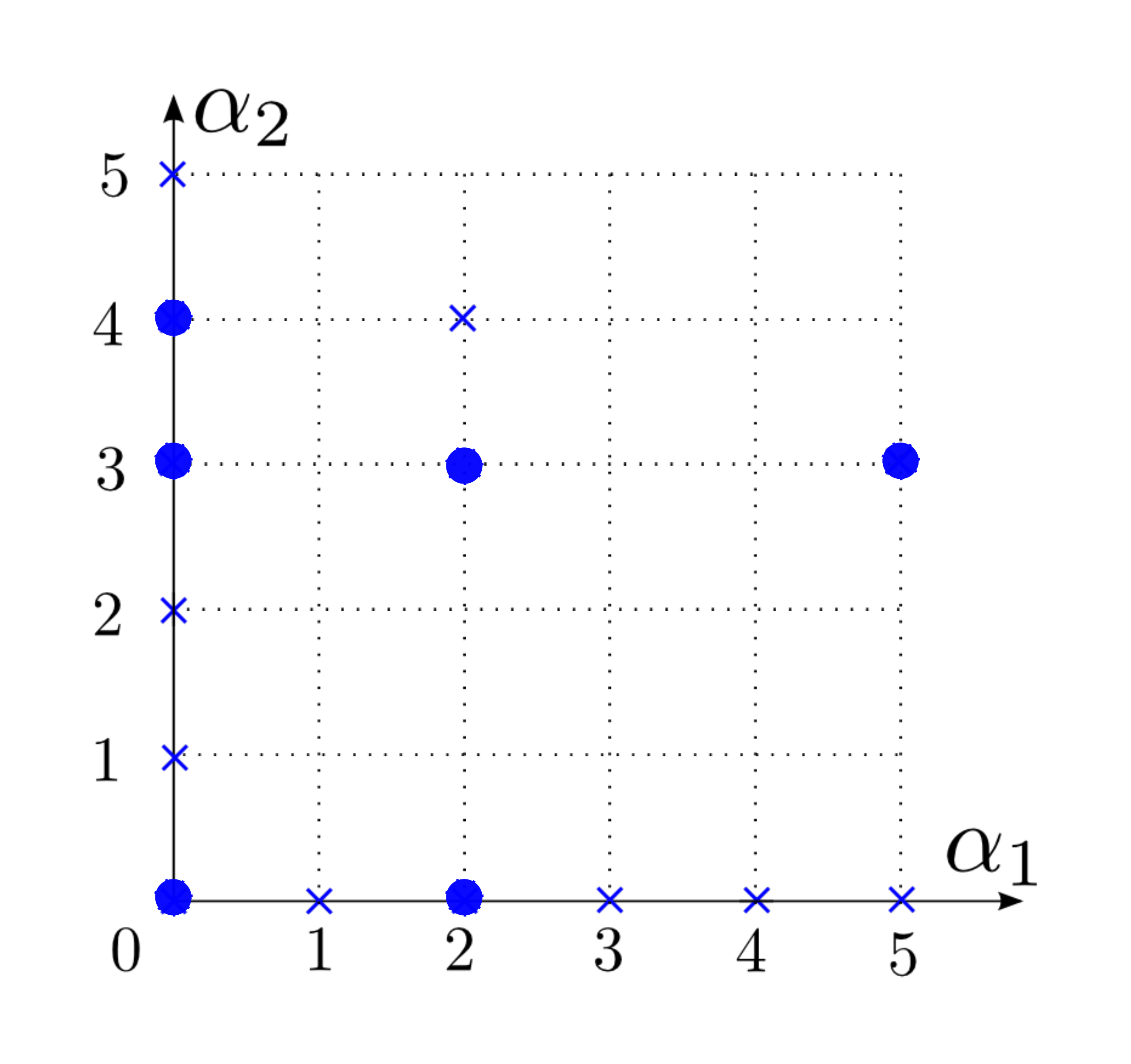}
    }
  \subfigure
  [Iteration 6b: weak heredity]
  	{
    \includegraphics[width=0.3\linewidth]{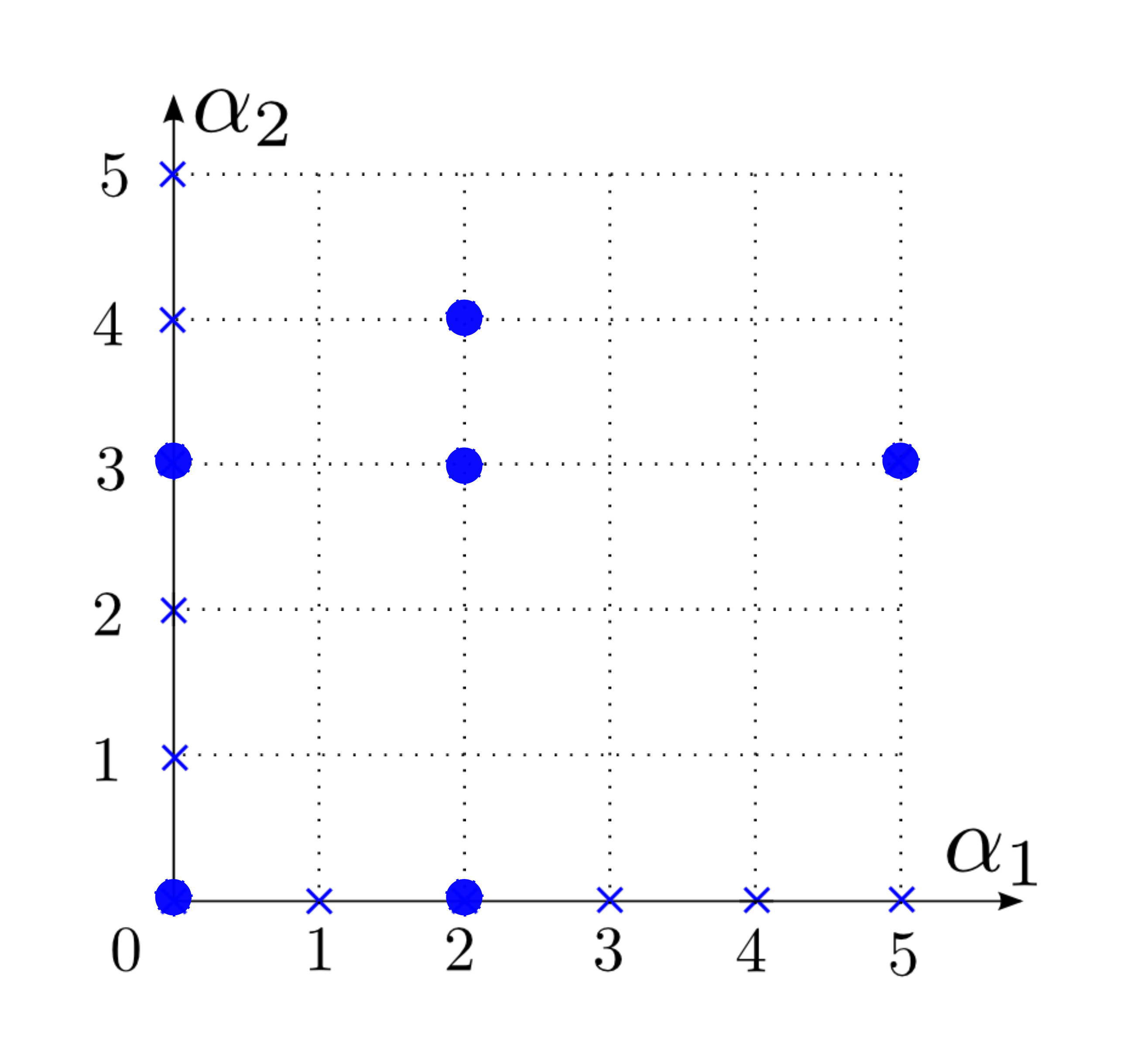}
    }
    \caption{Hierarchical adaptive PCE based on the heredity principle in the multi-index space. $p=5$ and $r=2$ are used for the truncation scheme. Second-rank PCE are assumed to be sufficiently accurate. The cross symbols (x) represent the candidate basis. The dot symbols ($\bullet$) represent the basis selected in the current step.}
    \label{fig3.2.1}
  \end{center}
\end{figure}

An iterative approach, named hierarchical adaptive PCE, is proposed to incorporate both forms of heredity in the LAR scheme. After detecting the most relevant term in the current iteration, the heredity principle is used for updating the current candidate set (which is used for later iterations). The proposed scheme works as follows:
\begin{enumerate}
\item Generate an initial candidate set, which consists of one-dimensional (1-D) polynomial chaos basis
\item For each iteration, perform the following steps until the stop condition is satisfied:
	\begin{enumerate}
		\item Find the most relevant term in the current candidate set.
		\item If it is a 1-D term
			\begin{enumerate}
				\item Generate interaction terms between the 1-D term selected in this iteration and the remaining 1-D terms in the current candidate set.
				\item Compare the relevance of the selected 1-D term with the generated interaction terms. The relevance of a basis element is represented by the correlation coefficient between it and the current residual. If the selected 1-D term remains the most relevant, then:
				\begin{itemize}
					\item Compute the associated coefficients as in LAR
					\item Update the candidate set by adding the interaction terms generated in the current iteration.
				\end{itemize}
				\item otherwise, select the most relevant interaction term, compute the associated coefficients as in LAR
			\end{enumerate}
		\item otherwise, it is an interaction term. Compute the associated coefficients as in LAR.
		\item Extract the selected term from the candidate set. Update the residual. Check the stop condition.
	\end{enumerate}
\end{enumerate}
The approach is illustrated in \figref{fig3.2.1} until six terms are active in the expansion. In iteration 4, the interaction term $(\alpha_1,\alpha_2)=(2,3)$ is selected after its parent terms $(2,0)$ and $(0,3)$ have been selected in the previous iterations. Thus, the selection in the current iteration follows the strong heredity principle. In iteration 5, the 1-D term $(5,0)$ is selected first, but it is less relevant than its child interaction term $(5,3)$, which is generated on-the-fly. The latter is therefore selected instead of its parent term. Only one parent of this interaction term, \ie $(\alpha_1,\alpha_2)=(0,3)$, is currently active. This selection illustrates the weak heredity principle. In iteration 6, the weak heredity prevails again when the interaction term $(2,4)$ is selected instead of its parent $(0,4)$.


Now we explain the superiority of the proposed approach compared to \cite{Yuan2007}, in which authors proposed distinct algorithms for strong and weak heredity.
If only the strong heredity principle is used, then 1-D terms must always be selected before their interaction terms. This may introduce error, when the interaction effects are important \cite{Shah2012}. If one uses only the weak heredity (which already covers the strong heredity), the interaction terms would quickly include the full candidate basis. In addition, in practical problems, it is impossible to know in advance whether the considered model follows the heredity principle or not, and if it does, in which form. In particular, a group of variables may follow the strong form whereas another group the weak. Our approach is flexible in the sense that both forms of heredity principle can be considered simultaneously in each iteration. 

It is worth to emphasize that one can also apply the hyperbolic and low-rank truncation schemes when generating the interaction terms according to the heredity principle. This helps reducing further the size of the candidate set. In this case, the proposed approach can be considered as an extension of the reference approach except that the candidate polynomial chaos basis is updated iteratively by means of the principle of heredity. The size of the candidate set in the proposed algorithm increases iteratively, while remaining always smaller than that of the reference approach. Therefore, it is expected to outperform the the reference approach, in particular when the considered model requires anisotropy in the polynomial basis, \ie high polynomial degrees in certain directions versus low degrees in others.


As a summary, this approach allows one to update the candidate basis in each iteration using the heredity principle rather than using a fix candidate set. One can discard a large number of interaction terms that are not relevant and just focus on the significant ones that are selected on-the-fly.


\section{NUMERICAL EXAMPLES}

The proposed hierarchical adaptive PCE (hereafter denoted by h-LAR) is implemented in UQLab \cite{MarelliICVRAM2014} (the Uncertainty Quantification toolbox in Matlab) developed at
the Chair of Risk, Safety \& Uncertainty Quantification (ETH Z\"{u}rich, Switzerland), which already includes the reference approach (hereafter denoted by LAR).

Let us examine the effectiveness of h-LAR with respect to LAR by means of two numerical examples.

\subsection{Sobol' function}

We consider the Sobol' function \cite{BlatmanThesis}:
\begin{eqnarray}
 Y= \prod\limits_{i=1}^8 \dfrac{\abs{4 X_i - 2} + c_i}{1+c_i}
\end{eqnarray}
in which $c =\acc{1,2,5,10,20,50,100,500}\tr$ and the input parameters are uniform random variables $X_i \sim \cu [0,1], \, i=1 \enum 8$. The Sobol' function is highly anisotropic in that it requires high polynomial degree for low-$i$ dimensions.

\begin{figure}[ht]
  \begin{center}
    \includegraphics[width=0.5\linewidth]{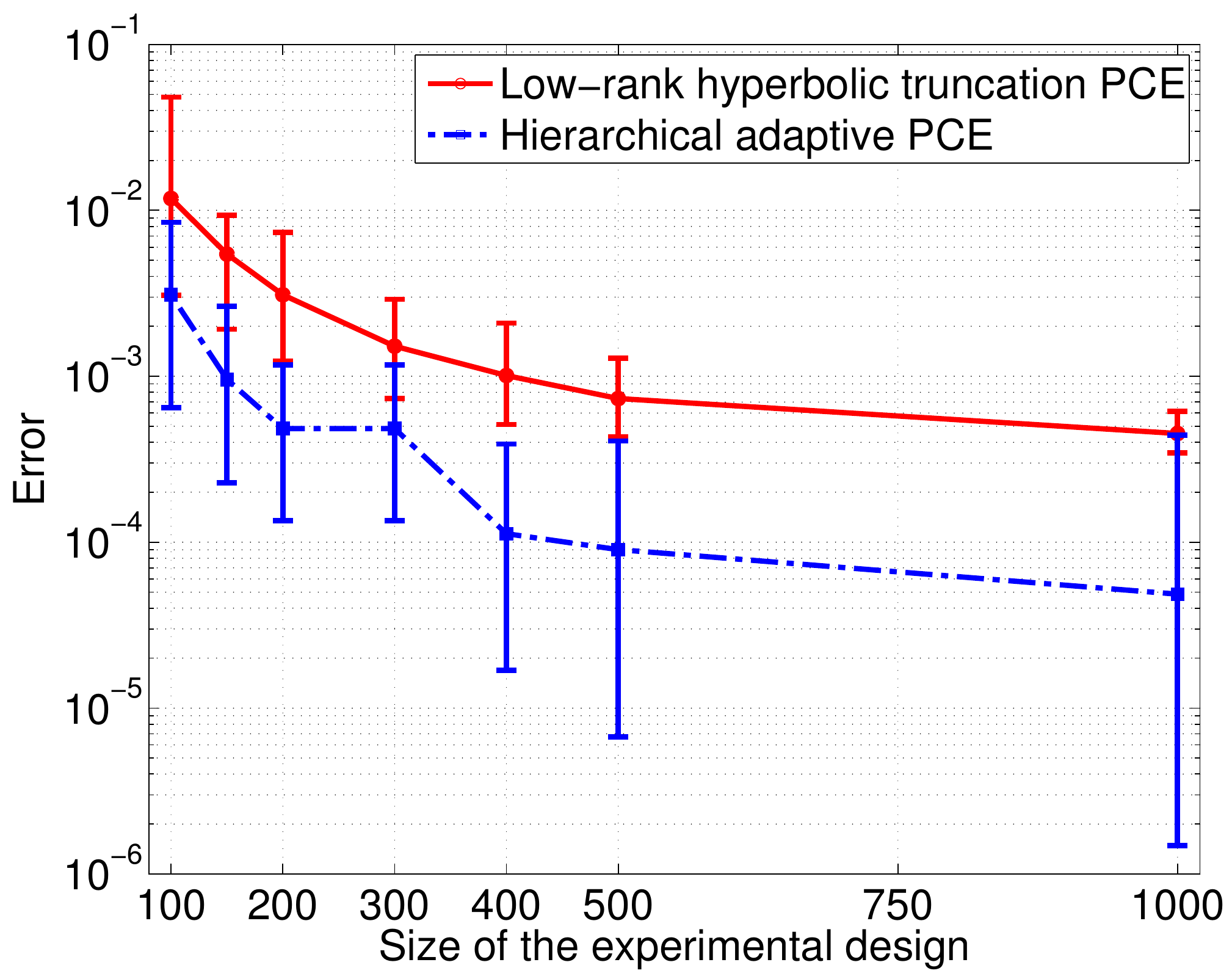}
    \caption{ Mean value and 95\% confidence interval of the leave-one-out error as a function of the experimental design's size. The mean values and confidence intervals are obtained with 100 Monte Carlo realizations.}
    \label{figSobol1}
  \end{center}
\end{figure}

We first investigate the effect of the experimental design's size on the accuracy of the resulting PCE models.
For a given size (\eg $100, 150, 200, 300, 400, 500, 1000$), one repeats the Latin hypercube sampling (LHS) of the experimental design 100 times then builds the corresponding PCE by means of LAR and h-LAR. The mean value and 95\% confidence interval of the estimated LOO errors are depicted in \figref{figSobol1}. Note that LOO error is chosen as an indicator of accuracy because it is much more sensitive to overfitting than the commonly used empirical error \cite{SudretBook2015}. h-LAR performs significantly better than LAR with faster convergence rate and smaller variability due to LHS. 

\begin{figure}[ht]
  \begin{center}
  \subfigure
    [Low-rank hyperbolic truncation scheme]
    	{
    \includegraphics[width=0.49\linewidth, height = 6 cm]{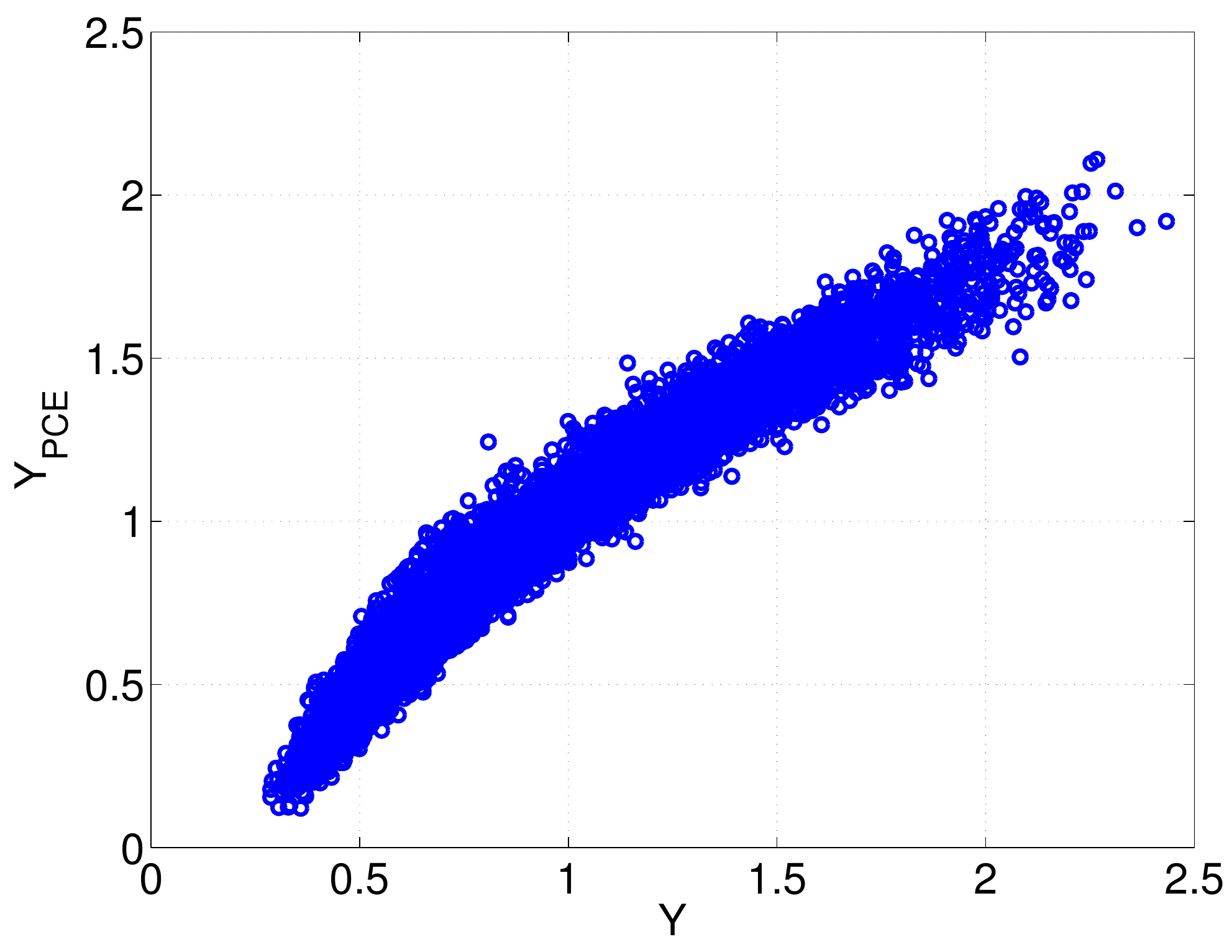}
    \includegraphics[width=0.49\linewidth, height = 6 cm]{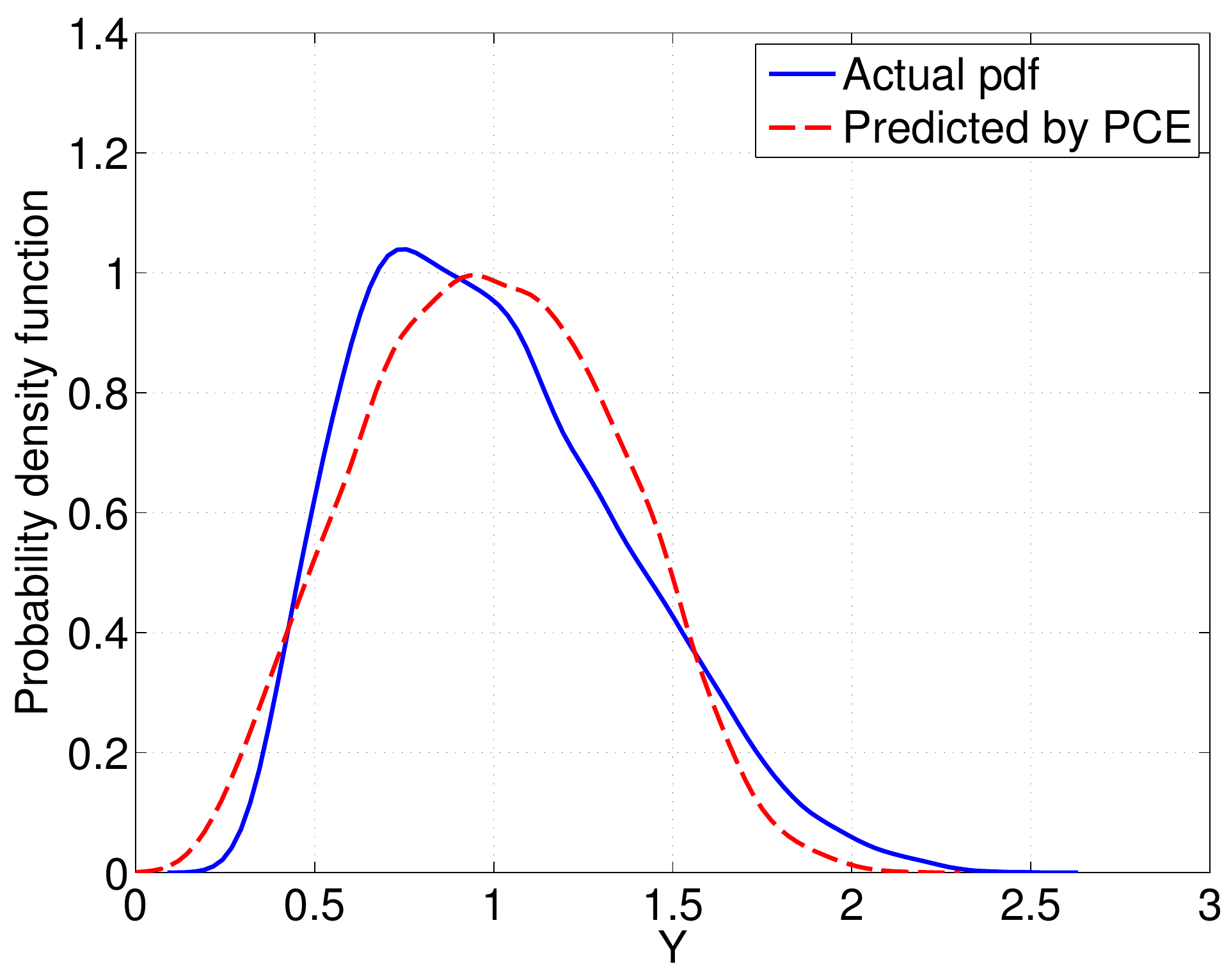}
    }
  \subfigure
  [Hierarchical adaptive PCE]
    {
    \includegraphics[width=0.49\linewidth, height = 6 cm]{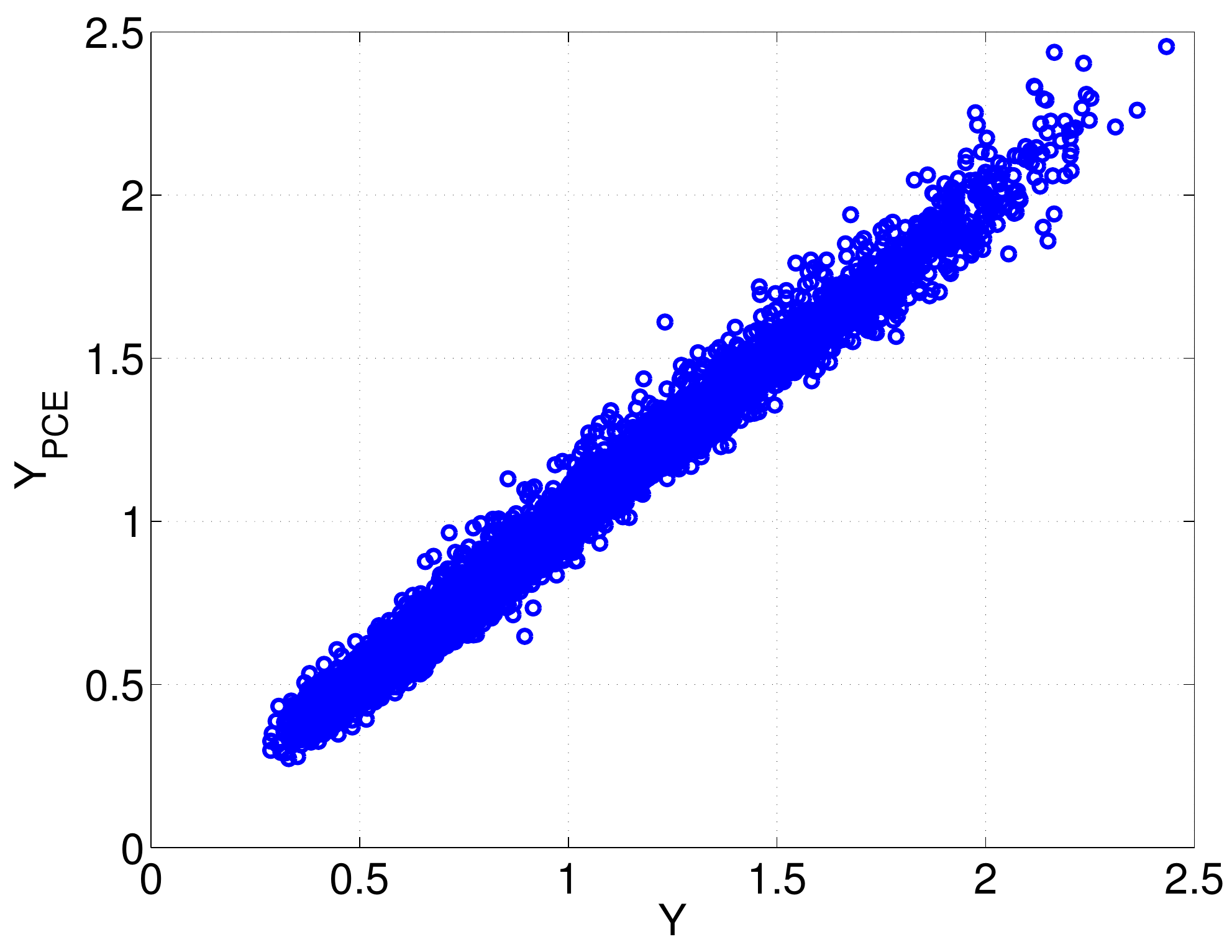}
    \includegraphics[width=0.49\linewidth, height = 6cm]{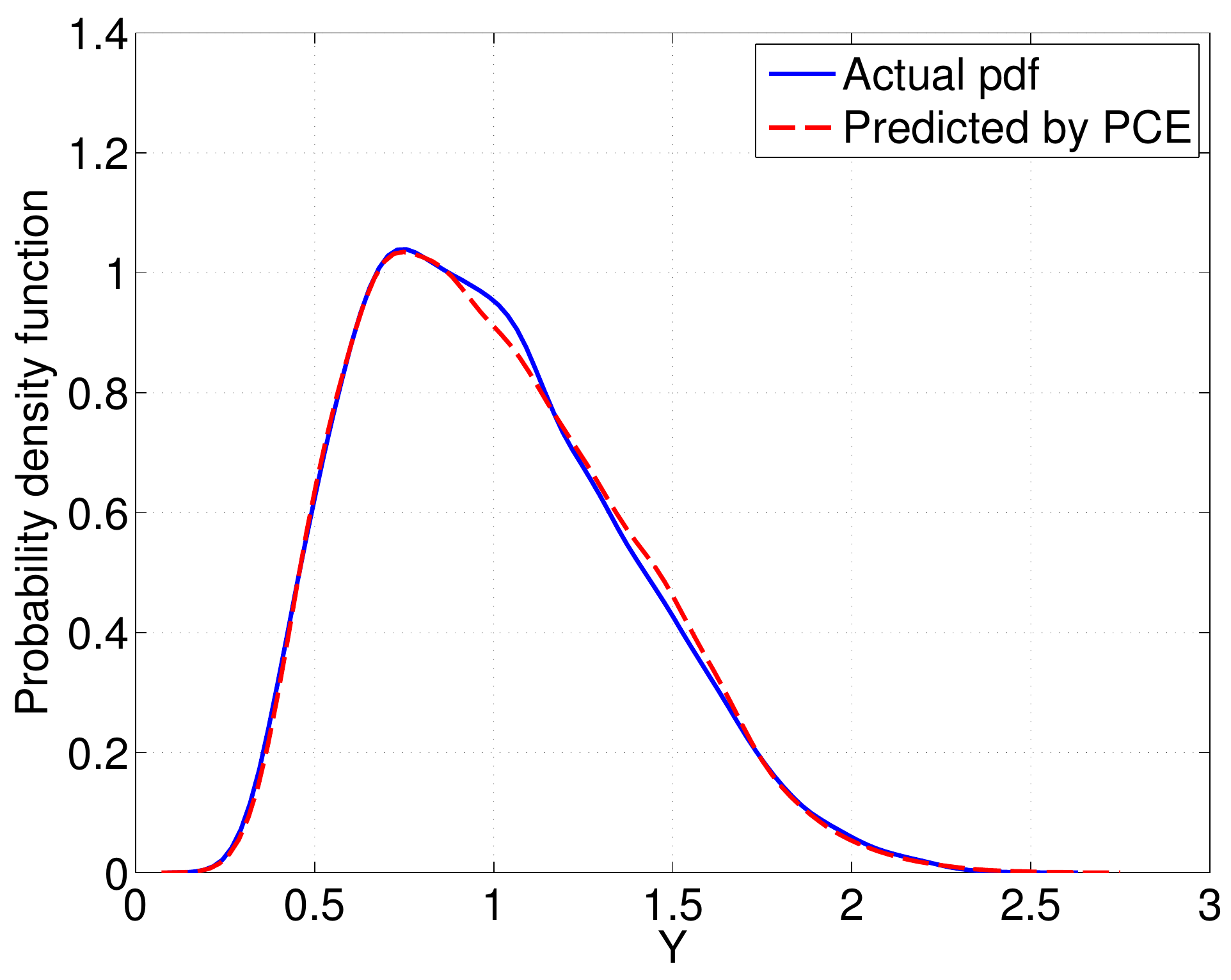}
    }
    \caption{Sobol' function - Experimental design of size 200 - PDF of the output.}
    \label{figSobol2}
  \end{center}
\end{figure}

\begin{figure}[ht]
  \begin{center}
  \subfigure
      [Low-rank hyperbolic truncation scheme]
      	{
    \includegraphics[width=0.49\linewidth, height = 6 cm]{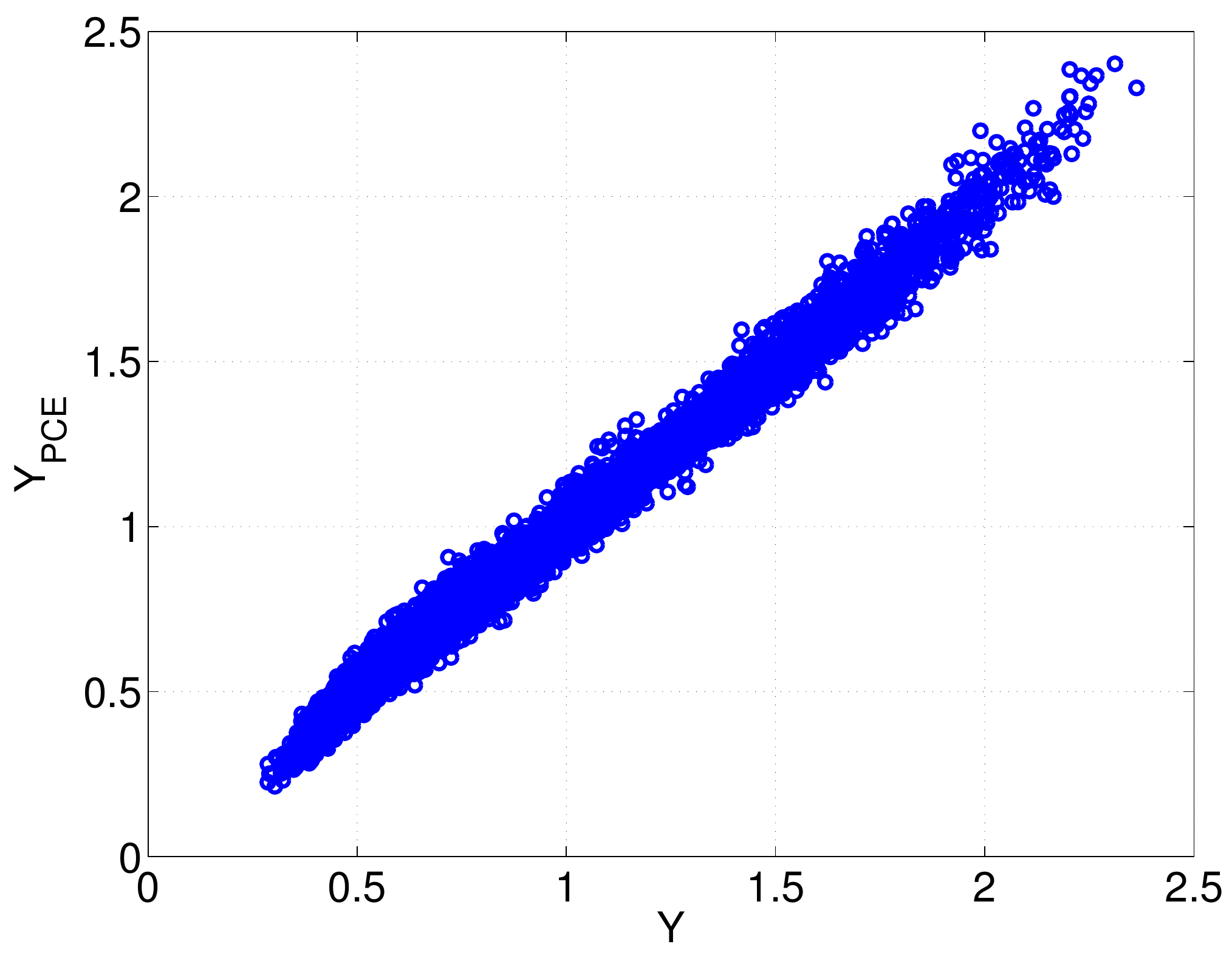}
    \includegraphics[width=0.49\linewidth,height = 6 cm ]{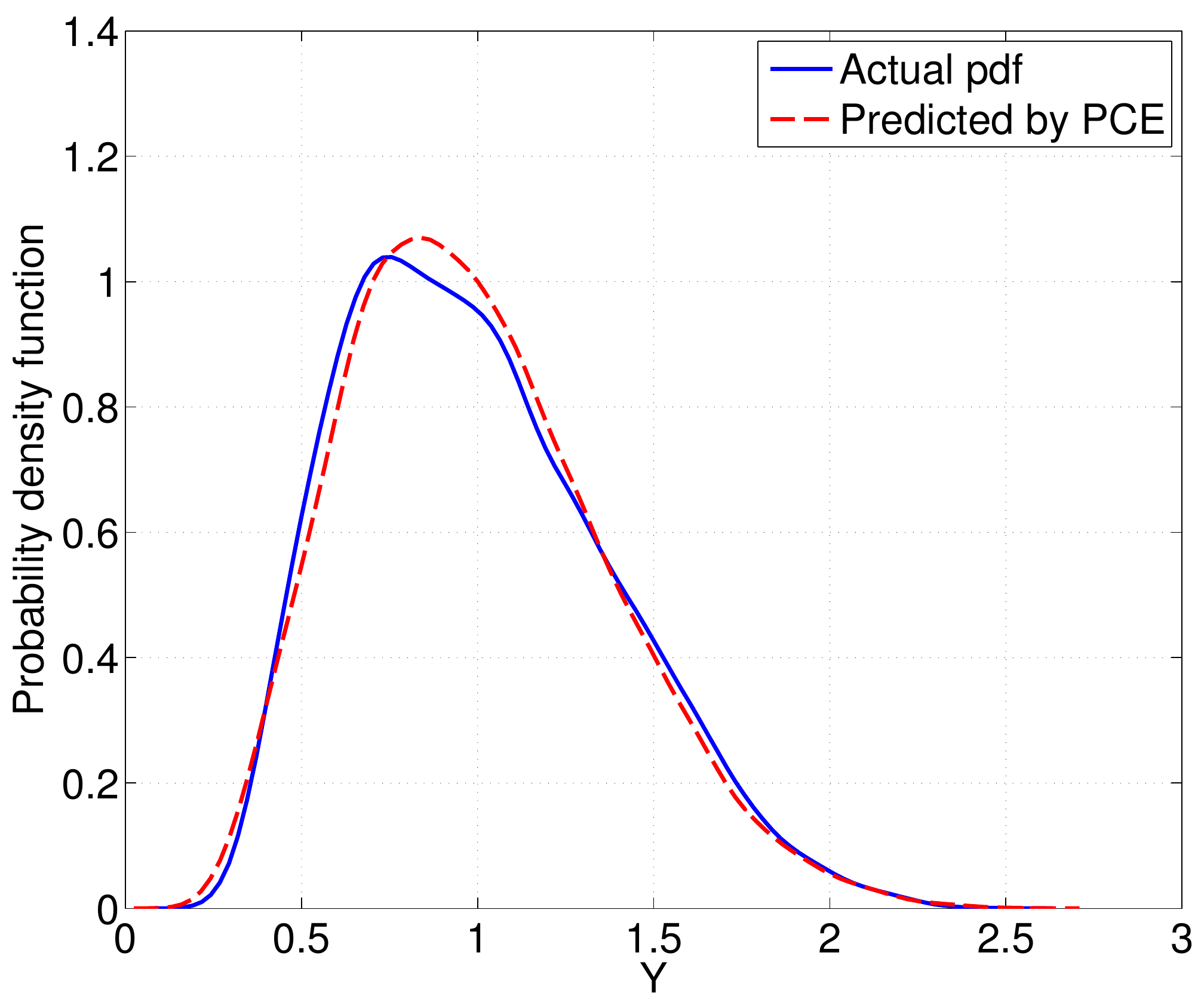}
    }
      \subfigure
      [Hierarchical adaptive PCE]
        {
    \includegraphics[width=0.49\linewidth, height = 6 cm]{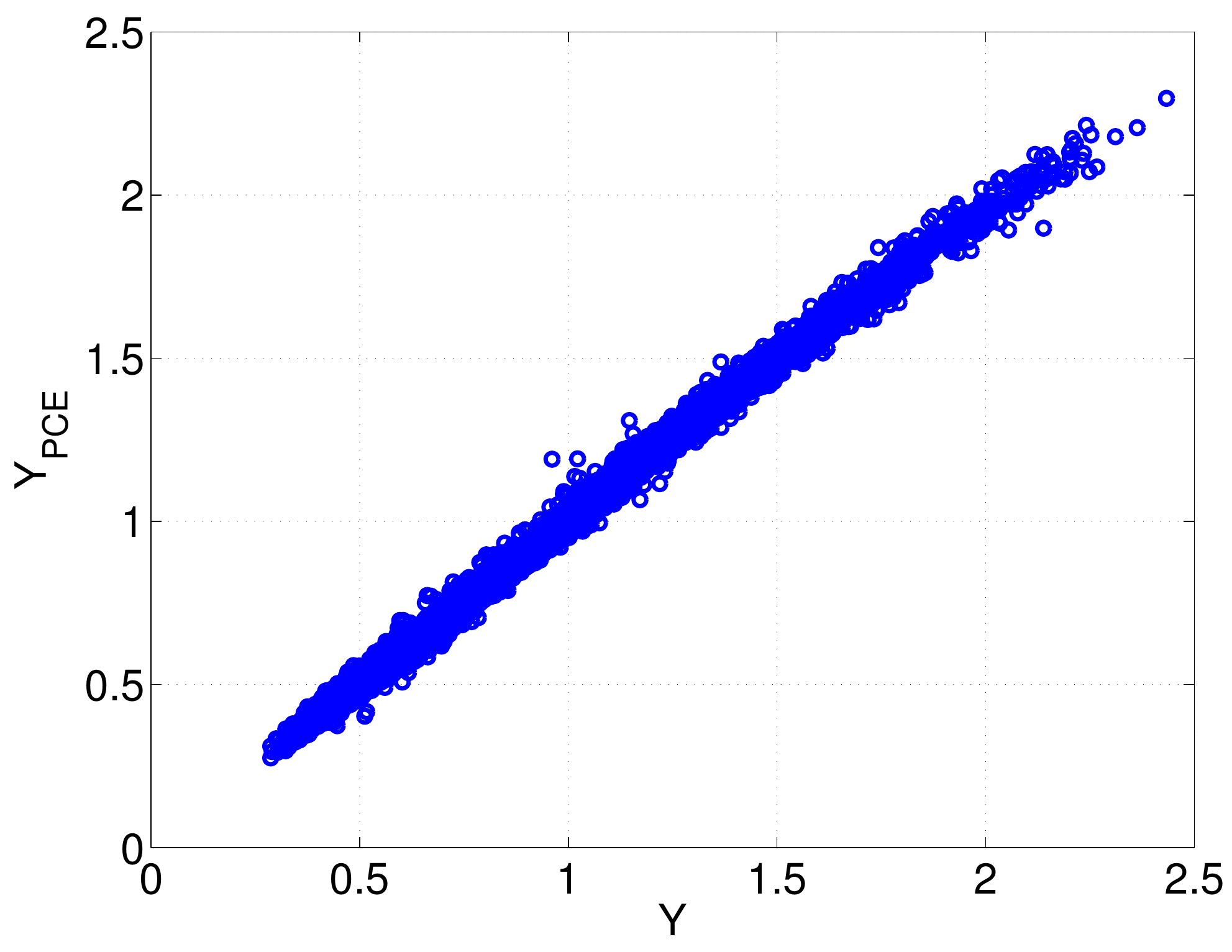}
    \includegraphics[width=0.49\linewidth, height = 6 cm]{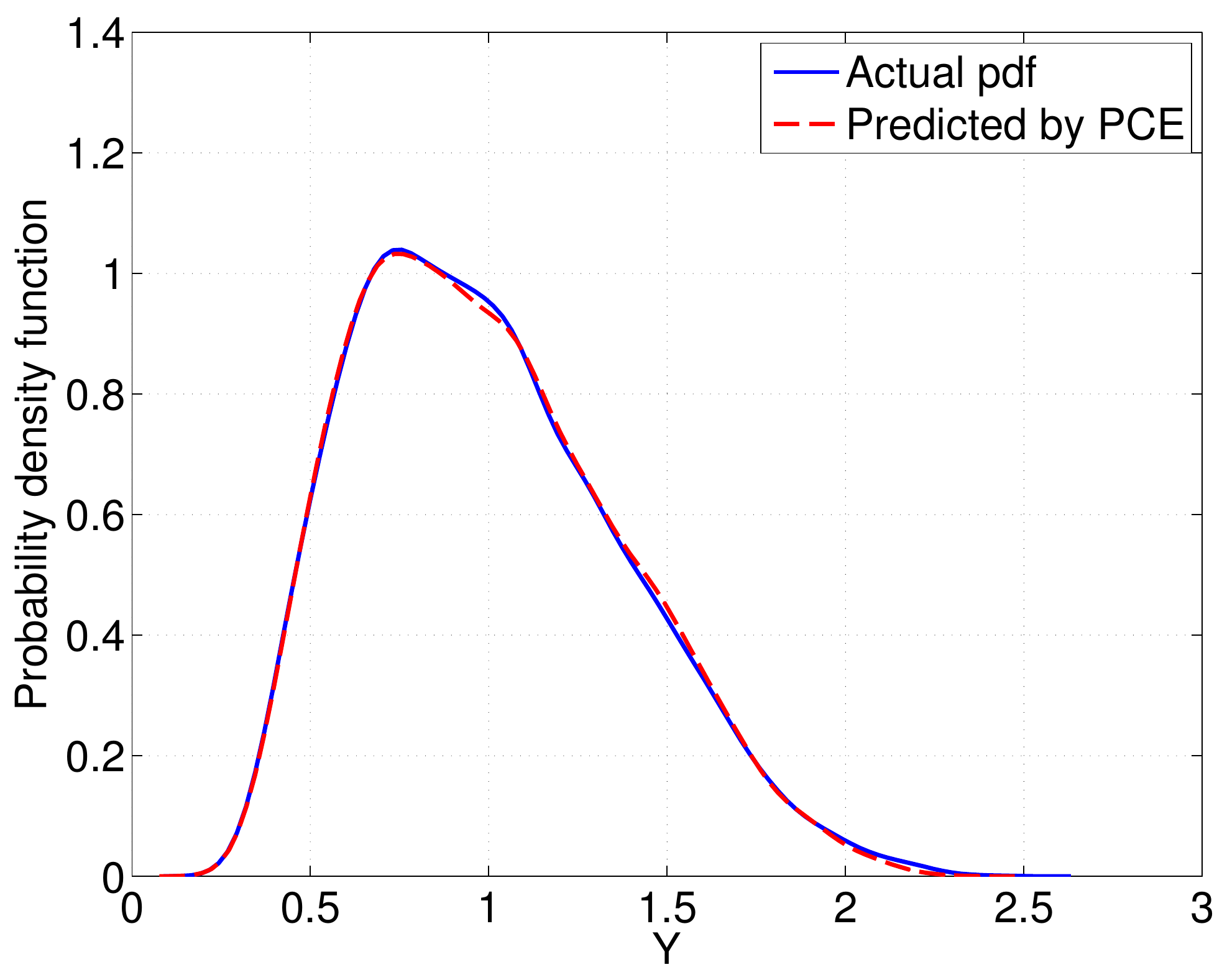}
    }
    \caption{Sobol' function - Experimental design of size 300 - PDF of the output.}
    \label{figSobol3}
  \end{center}
\end{figure}

As a further validation, given an ED of size 200, we compute the PCE by the two approaches using the same truncation options ($p=30$, $q=0.5$, $r=2$). Using the computed PCE, we predict the output values on a new set of input parameters $\acc{\ve{\cx_i}, i=1 \enum 10^5}$. The predicted values of the output and its probability density function (PDF) are compared with the values and the PDF computed with the analytical model. The results are shown in \figref{figSobol2}. Similarly, \figref{figSobol3} depicts the results when an ED of size 300 is used.
The corresponding results (\eg polynomial degree giving the best accuracy, relative validation error $\epsilon_{V}$ and number of retained polynomials $N_r$) of the PCE are given in Table~\ref{tab:2}.
In both cases, h-LAR outperforms LAR in predicting specific values of the output as well as its statistical distribution.  One sees that given the same computational budget (\ie the same ED), h-LAR allows one to reduce the validation error by a factor of 5. The best degree in h-LAR is higher than that in LAR, which leads to a larger number of selected terms in the expansion.

\begin{table}[ht!]
\centering
\begin{tabular}{lllllll}
\hline
 Method & Size of ED & $q$ & $r$ & Best degree & $\epsilon_V$ & $N_r$ \\
\hline
 LAR & 200 & 0.5 & 2 & 7 & $5.95 \times 10^{-2}$ & 37 \\
 h-LAR & 200 & 0.5 & 2 & 9 & $1.6 \times 10^{-2}$ & 100 \\
 LAR & 300 & 0.5 & 2 & 9 & $2.2 \times 10^{-2}$ & 88 \\
 h-LAR & 300 & 0.5 & 2 & 20 & $4.3 \times 10^{-3}$ & 154 \\
\hline
\end{tabular}
\caption{Options and results of the utilized PCE}
\label{tab:2}
\end{table}

\subsection{Schwefel function}

We now consider a modified version of the Schwefel function \cite{Laguna2005} which reads:

\begin{eqnarray}
\begin{split}
 Y = & - \sum\limits_{i=1}^{20} (\frac{i}{20}+0.5) X_i \, \sin(\sqrt{ \frac{i}{20} \, \abs{X_i} }) \\
          &   + \frac{1}{3000} \sum\limits_{j\in\cs_1} (\frac{j}{20}+0.5) X_j \, \sin(\sqrt{ \frac{j}{20} \, \abs{X_j} }) \, \sum\limits_{k\in\cs_2} X_k \\
\end{split}
\end{eqnarray}
where $X_i \sim U[-500,500], \, i=1 \enum 20$, $\cs_1 = \acc{1, 3, 5, 6, 8}$ and $\cs_2 = \acc{15, 18, 20}$. This is a complex function with many local minima.
The first term is modified in such a way that the function requires high polynomial degree in high-$i$ dimensions and the second term is added to the function in order to introduce interaction effects between some of the dimensions.

Given an ED of size $1,000$, we compute the PCE of the Schwefel function by means of the two approaches with the same truncation options, \ie $p=30$, $q=0.25$, $r=2$. 
The values of the function on $10^5$ samples of the input parameters are predicted by the two PCE models and compared with the values obtained with the numerical functions. 
The outcomes of the PCE are presented in Table~\ref{tab4.2.1}. The relative validation error of h-LAR is smaller than that of LAR, whereas more polynomials are retained by h-LAR. For this numerical model, h-LAR performs slightly better than LAR.

\begin{table}[ht!]
\centering
\begin{tabular}{lllllll}
\hline
 Method & Size of ED & $q$ & $r$ & Best degree & $\epsilon_V$ & $N_r$ \\
\hline
 LAR & 1000 & 0.25 & 2 & 23 & $2.2 \times 10^{-2}$ & 556 \\
 h-LAR & 1000 & 0.25 & 2 & 30 & $1.03 \times 10^{-2}$ & 671 \\
\hline
\end{tabular}
\caption{Options and results of the utilized PCE}
\label{tab4.2.1}
\end{table}


%

\section{CONCLUSIONS AND PERSPECTIVES}

In modern engineering and computational sciences, polynomial chaos expansions (PCE) have been widely used as a powerful tool for uncertainty quantification. The application of PCE to high-dimensional complex problems might be, however, hindered due to computational limitations. The reference sparse PCE technique relies on selecting the relevant polynomial basis functions from a candidate set defined \emph{a priori} by means of appropriate truncation schemes. Recent research shows that it can be more effective to compute PCE with an adaptive basis enrichment approach instead. 

In the current paper, we propose a novel hierarchical adaptive-basis approach, which consists in selecting the important basis elements from an iteratively enriched candidate basis. The proposed approach is based on combining the least-angle-regression method with the principle of heredity. The latter, which originated in a biological context, is herein utilized in a statistical sense for detecting the potentially relevant interaction effects in the model. Using some numerical models, we proved the effectiveness of our approach.

Further investigations are required in order to take into account higher interaction order. In addition, the current approach updates the candidate basis set only by adding new terms, while it can also be important to remove irrelevant terms during the process, which may contribute to further decreasing the computational costs and increasing the accuracy of PCE.

\section{Acknowledgment}
The first author would like to thank Dr. Stefano Marelli (ETH Z\"{u}rich, Chair of Risk, Safety \& Uncertainty Quantification) for various fruitful discussions during the preparation of the paper. We also acknowledge the Development team of UQLab (the Uncertainty Quantification toolbox in Matlab) for providing the numerical platform on which this work is carried out.





\end{document}